\documentclass[%
 reprint,
 amsmath,amssymb,
 aps,
]{revtex4-2}

\usepackage{graphicx}
\usepackage{dcolumn}
\usepackage{bm}

\usepackage{subcaption}
\usepackage{amsmath,amssymb,amsfonts}
\usepackage{algorithmic}
\usepackage{graphicx}
\usepackage{textcomp}
\usepackage{physics}
\usepackage{xcolor}
\usepackage{ulem}
\usepackage{graphicx}
\newcounter{algorithmnumber}

\newcommand{\etal}{\textit{et al.}}

\begin{document}

\preprint{APS/123-QED}
\title{A Neutral Atom-Based Hybrid Classical-Quantum Approach for the Entanglement Routing Problem}

\author{M. Yassine Naghmouchi}
\email{yassine.naghmouchi@pasqal.com}
\affiliation{Pasqal, 24 rue Emile Baudot, 91120 Palaiseau, France}

\author{Quentin Ma}
\affiliation{Pasqal, 24 rue Emile Baudot, 91120 Palaiseau, France}

\author{Agathe Blaise}
\affiliation{Thales SIX GTS France, 92230 Gennevilliers, France}

\author{René Veyland}
\affiliation{Thales SIX GTS France, 92230 Gennevilliers, France}

\author{Wesley Coelho}
\affiliation{Pasqal, 24 rue Emile Baudot, 91120 Palaiseau, France}

\date{\today}

\begin{abstract}
The efficient distribution of end-to-end entanglement is a key requirement for enabling distributed quantum communication and computing applications. Quantum Information Networks (QINs) therefore require efficient routing mechanisms under limited resources and fidelity constraints. We study the Entanglement Routing in Quantum Networks (ERQN) problem, formulated as a fidelity-constrained unsplittable multicommodity flow problem that maximizes the number of admitted entanglement requests. As a proof of concept, we investigate the integration of neutral-atom quantum optimization into a hybrid classical--quantum column generation framework. This study does not claim quantum utility, or scalability. Instead, it provides an initial assessment of whether a neutral-atom-based routine can generate useful routes relative to a standard classical heuristic, thereby examining its potential role within hybrid decomposition frameworks. The framework combines a classical restricted master problem, which selects routes among the available candidates, with a pricing problem that generates new fidelity-feasible paths. The pricing problem is an NP-hard constrained shortest-path problem formulated as a Quadratic Unconstrained Binary Optimization (QUBO) problem. The resulting QUBO is addressed using a neutral-atom quantum workflow combining hardware-aware register embedding and instance-driven pulse shaping. To the best of our knowledge, this is the first investigation of a neutral-atom quantum optimization routine as a pricing oracle for fidelity-constrained entanglement routing. Experiments conducted using neutral-atom quantum processor emulators on small yet representative benchmark instances show that, when combined with warm-start and post-processing, the proposed method achieves an optimality gap below $1\%$ across all tested sizes. The selected classical counterpart, based on simulated annealing for route generation, exhibits gaps of up to $6\%$. The results suggest that the quantum-generated bitstrings yield routes that are more amenable to classical refinement within the column generation framework. Overall, this study demonstrates the feasibility of the proposed integration under the considered experimental setting and provides an encouraging initial indication that neutral-atom quantum routines can contribute useful candidate solutions within hybrid optimization frameworks. It also motivates further evaluation on larger instances, stronger classical baselines, and neutral-atom quantum hardware.
\end{abstract}


\maketitle


\section{Introduction}
\label{sec:introduction}

\begin{figure}[!t]
\centering
\includegraphics[width=\linewidth]{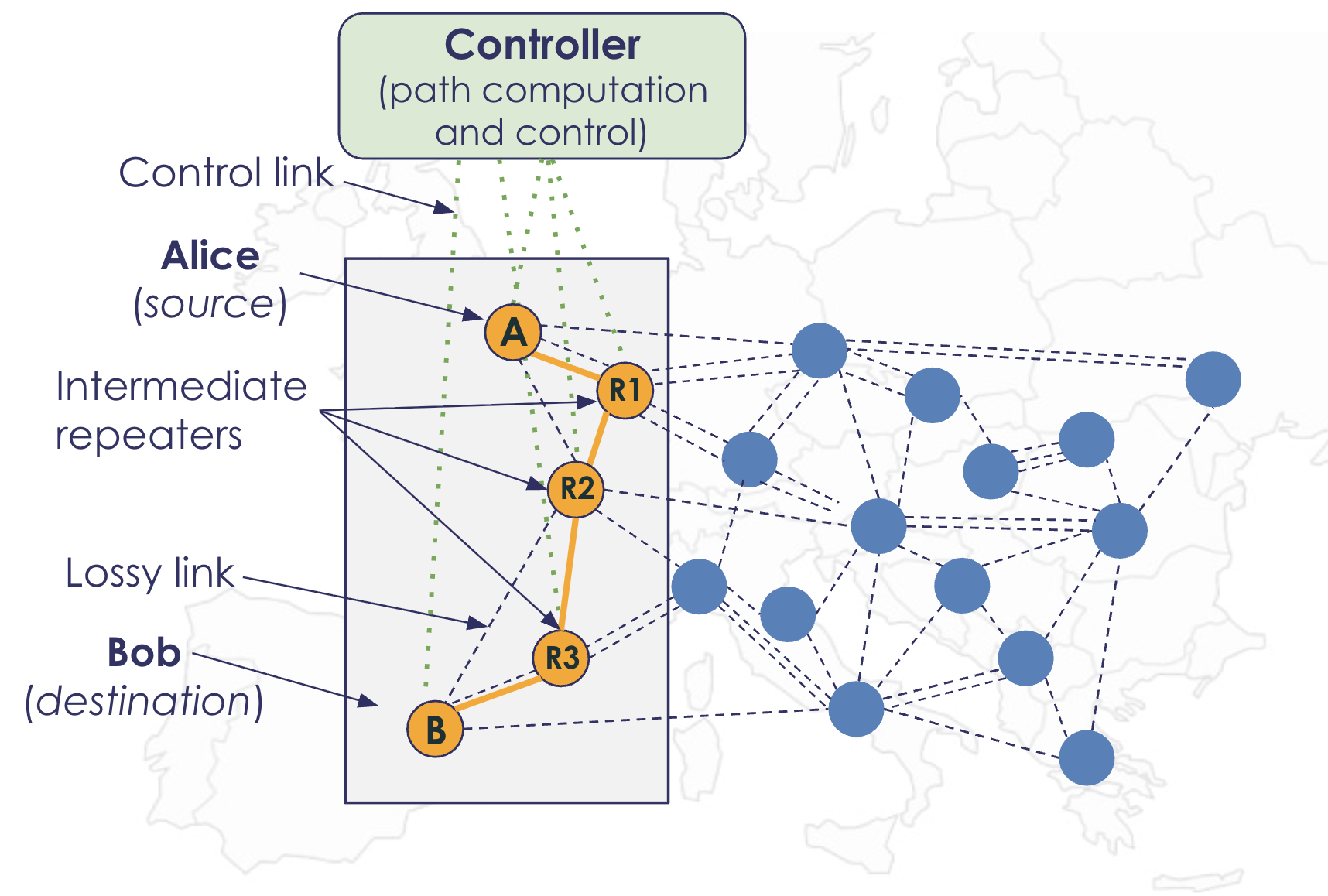}
\caption{Quantum network architecture for entanglement routing. A centralized controller monitors stochastic link availability and computes routing decisions each generation round via classical control links. The local segment shows Alice (\textit{source}) and Bob (\textit{destination}) connected through three intermediate repeaters R1, R2, R3 over lossy quantum links. The orange mesh illustrates the quantum backbone over which E2E entanglement is established via entanglement swapping at intermediate repeaters.}
\label{fig:quantum_net2}
\end{figure}

Quantum Information Networks (QINs) aim to distribute entanglement over long distances, enabling the interconnection of remote quantum systems~\cite{wehner2018quantum_internet}. Such networks underpin transformative applications, including quantum-secure communications, distributed quantum sensing, and delegated quantum computation. The fundamental resource of a QIN is end-to-end (E2E) entanglement. Unlike classical signals, quantum signals cannot be amplified, so long-distance transmission relies on \textit{quantum repeaters}~\cite{briegel1998quantum,sangouard2011quantum_repeaters_review} and \textit{entanglement swapping}~\cite{zukowski1993entanglement_swapping,pan1998experimental_entanglement_swapping}. In a simple three-node setting, two neighboring links first establish entangled Einstein--Podolsky--Rosen (EPR) pairs, as highlighted in Figure~\ref{fig:quantum_net2}. An intermediate repeater performs a joint measurement, after which classical communication enables the endpoints to share E2E entanglement and perform quantum teleportation.

A central challenge of QINs is that local link generation is probabilistic and may fail with non-negligible probability. In practice, these outcomes are heralded and reported to a centralized controller, which therefore has real-time knowledge of which elementary links are currently available. Because this availability changes from one generation round to the next, routing decisions must be made rapidly over an ephemeral and stochastic network state. In this work, we focus on the per-round routing decision after the outcomes of elementary-link generation have been heralded and communicated to the controller. Conditioned on this observed network snapshot, the available links, their capacities, and the fidelity-degradation factors are treated as known inputs, yielding a deterministic optimization problem; link-generation and swapping success probabilities, waiting times, and finite entanglement lifetimes are therefore not explicitly modeled in this work.
In this context, routing consists in selecting chains of available elementary links and intermediate repeaters along which entanglement swapping can establish E2E entanglement for the requested source-destination pairs. This leads to the \textit{entanglement routing} problem~\cite{Meter2021, gyongyosi2017entanglement}: at each generation round, the controller must determine how to route a set of connection requests over the currently available links while respecting capacity and fidelity constraints, before the network state changes again. Recent surveys have emphasized that entanglement routing differs from classical routing not only because of network-state information, but also because routing decisions must account for fidelity requirements, probabilistic entanglement generation and swapping, and the finite lifetime of entangled states~\cite{abane2025entanglement}. In this taxonomy, routing schemes are commonly classified according to when paths are computed with respect to entanglement generation, leading to proactive, reactive, virtual, and opportunistic routing families.

This routing problem can be naturally viewed in a dynamic or online setting. In practice, entanglement requests arrive over time, forming a batch of active requests to be served at each time slot. At each such time slot, the resulting routing task can be formulated as a variant of the Unsplittable Multicommodity Flow (UMCF) problem~\cite{gamst2012comparing}, augmented with fidelity constraints. Each connection request is characterized by a source-destination pair, a given entanglement generation rate (number of channels), and a minimum acceptable end-to-end fidelity. Each request must be routed along a single path---thereby enforcing unsplittable routing---while satisfying link-capacity constraints and an end-to-end fidelity threshold that degrades exponentially with the number of hops due to decoherence and imperfect entanglement swapping. The controller must therefore decide, for the current batch of requests, which ones to admit and how to route them so as to maximize the aggregate served demand while satisfying all capacity and fidelity constraints. The resulting optimization problem is NP-hard~\cite{even1976complexity}.

Existing solution approaches typically fall into two broad categories. Fast heuristic or approximation methods can provide feasible solutions at moderate computational cost, but often with limited guarantees on solution quality~\cite{halder2024optimal,nguyen2025maximizing}. Exact optimization approaches based on integer programming can in principle provide proven optimal solutions when solved to completion, but they often become computationally prohibitive as the network size and the number of requests increase~\cite{halder2024optimal,nguyen2025maximizing,zeng2023entanglement}. While the former may lead to suboptimal solutions, the latter typically suffers from limited scalability.

Motivated by the benefit of achieving this trade-off between scalability and solution quality, we
investigate the potential of a \textit{hybrid classical--quantum column generation} framework designed to balance scalability and solution quality. The idea is to avoid solving the full problem monolithically, as in exact integer-programming approaches, while retaining a stronger optimization structure than purely heuristic methods. Column generation is a well-established decomposition technique for large-scale linear and integer programming problems~\cite{Desrosiers2006}. In our framework, it is assissted by neutral atom-based quantum subroutines. Under this approach, the problem is decomposed into two interacting components: the \textit{Restricted Master Problem} (RMP), which selects the best combination of routing configurations from a restricted subset of feasible paths while maximizing the number of admitted demands; and the \textit{Pricing Subproblems} (PSPs), which use the current RMP solution to generate new promising paths by identifying fidelity-compliant routes that can improve the overall objective value. This iterative process continues until no new path can further improve the objective function, as established by standard column-generation theory~\cite{Desrosiers2006}.

The pricing subproblems, which can be seen, in the studied problem, as a variant of the constrained shortest path problem~\cite{feng2014lagrangian}, help to solve several simpler but still difficult optimization problems. To address this challenging bottleneck of the decomposition method, we formulate the PSPs as Quadratic Unconstrained Binary Optimization (QUBO) instances and solve them using a neutral atoms-based quantum sampler that is specially tailored to encoding the QUBO cost function and generating efficient paths for each related entanglement demand. The proposed hybrid quantum-classical workflow combines the scalability of the column generation-based decomposition method with the sampling capabilities of neutral atom quantum processors to solve the Entanglement Routing problem.

The remainder of the paper is organized as follows. Section~\ref{sec:neutral_atoms} provides background on neutral atom quantum processors for combinatorial optimization. Section~\ref{sec:related_work} reviews related work on entanglement routing and quantum optimization approaches. Section~\ref{sec:problem_statement} formally defines the ERQN problem and introduces its mathematical formulation. Section~\ref{sec:hybrid_cg} presents the proposed hybrid classical--quantum column generation framework, including the restricted master problem, the pricing subproblem, its QUBO reformulation, and the design of the quantum routine. Section~\ref{sec:numerical_results} reports the numerical results. Finally, Section~\ref{sec:conclusion} concludes the paper and outlines future research directions.

\section{Neutral Atoms for Combinatorial Optimization}
\label{sec:neutral_atoms}

Combinatorial optimization problems arise in a wide range of industrial and scientific applications, from network routing~\cite{Toth2014, Ahuja1993} and scheduling~\cite{Pinedo2008} to resource allocation and logistics. Many of these problems belong to the NP-hard complexity class, meaning that finding optimal solutions becomes computationally intractable as the problem size grows~\cite{GareyJohnson1979}. Classical heuristics such as simulated annealing or greedy algorithms can provide approximate solutions, but often struggle to scale efficiently for large, complex instances~\cite{Lucas2014}.

Quantum computing has emerged as a promising paradigm for tackling such problems, with approaches including the Quantum Approximate Optimization Algorithm (QAOA)~\cite{Farhi2014} and quantum annealing~\cite{Farhi2000}. Among the various hardware platforms available in the current Noisy Intermediate-Scale Quantum (NISQ) era~\cite{Preskill2018}, neutral atom quantum processors based on Rydberg interactions have attracted considerable attention due to their long coherence times, and flexible qubit connectivity~\cite{Henriet2020, Browaeys2020}.

A distinctive feature of neutral atom platforms is the ability to reconfigure the geometry of the qubit register from shot to shot. Atoms trapped in programmable arrays of optical tweezers can be spatially arranged to directly embed the structure of a graph into the hardware~\cite{Ebadi2022}. When driven to highly excited Rydberg states, these atoms experience strong dipole-dipole interactions whose strength decays as $\sim 1/r^6$, where $r$ is the interatomic distance. The resulting Rydberg blockade mechanism, which prevents two nearby atoms from being simultaneously excited, naturally enforces an independent set constraint on the associated interaction graph~\cite{Lukin2001, Jaksch2000}.

The dynamics of such a system of interacting atoms is governed by the following Hamiltonian:
\begin{equation}
H(t)=\sum_i \frac{\Omega(t)}{2}\sigma_i^x -\sum_i \delta_i(t) n_i +\sum_{i<j} V_{ij} n_i n_j,
\end{equation}
where $\Omega(t)$ is the global Rabi frequency driving coherent transitions between the ground state $\ket{g}$ and the Rydberg state $\ket{r}$, $\delta_i(t)$ is a site-dependent detuning that can be individually addressed for each atom~$i$, $\sigma_i^x = \ket{g_i}\!\bra{r_i} + \ket{r_i}\!\bra{g_i}$ is the Pauli operator, $n_i = \ket{r_i}\!\bra{r_i}$ is the Rydberg state occupation operator, and $V_{ij}$ is the van der Waals interaction coefficient between atoms $i$ and~$j$, given by $V_{ij}=C_6/R_{ij}^{6}$, where $R_{ij}$ denotes their interatomic distance. The first term drives Rabi oscillations between the two atomic levels, the second term controls the energy landscape through locally tunable detunings, and the third term encodes the interaction graph through the spatial arrangement of the atoms.

This native mapping between the atomic register and graph structure makes neutral atom processors particularly well-suited for solving different classes of optimization problems. Indeed, once the register is prepared, the system is evolved by slowly sweeping the parameters $\Omega(t)$ and $\delta_i(t)$, for instance via a quantum adiabatic protocol, so that one can prepare the ground state in the regime $\Omega \to 0$, $\delta_i > 0$, which encodes the QUBO solution. The site-dependent nature of $\delta_i(t)$ further enables extensions to better encode complex cost functions~\cite{Dalyac2021}. Furthermore, recent advances in ancillary qubit encoding and graph reduction techniques have broadened the class of addressable problems beyond native graphs~\cite{Ebadi2022, Pichler2018}, allowing the encoding of arbitrary QUBO instances~\cite{Nguyen2023, Leclerc2024}.

In practice, leveraging neutral-atom processors for combinatorial optimization requires a workflow that connects problem formulation to physical execution. First, the target optimization problem is reformulated as a QUBO model, so that low-energy binary configurations correspond to high-quality candidate solutions. Second, the binary variables of the QUBO are mapped onto physical atoms through a register-embedding procedure, which determines atomic positions and therefore the effective interaction coefficients $V_{ij}$ induced by the geometry of the register. Third, a pulse sequence is designed by specifying the time-dependent controls $\Omega(t)$ and $\delta_i(t)$ so that the final Hamiltonian reproduces, as closely as possible, the energy landscape of the encoded QUBO while maintaining sufficient quantum exploration during the evolution. Finally, the resulting pulse is executed on the neutral-atom processor, and the measured bitstrings are interpreted as candidate solutions to the original optimization problem. This general workflow underlies the quantum pricing routine developed in this paper.

\section{Related Work}
\label{sec:related_work}

Entanglement routing has been extensively studied in the context of quantum repeaters and multi-hop entanglement distribution. A recent comprehensive survey organizes this literature by distinguishing the routing phase, which includes path computation and route installation, from the forwarding phase, which includes entanglement generation, swapping, purification, and reliability mechanisms~\cite{abane2025entanglement}. Within this taxonomy, routing schemes are commonly classified as proactive, reactive, virtual, or opportunistic, while routing algorithms include Dijkstra-based methods, graph path-search procedures, linear programming and multicommodity-flow formulations, greedy methods, and AI-based approaches~\cite{abane2025entanglement}. 
Our work falls within the path-computation component of this taxonomy. More specifically, we address an optimization fidelity-constrained routing problem for multiple concurrent entanglement requests, and focus on the algorithmic bottleneck associated with generating high-quality feasible paths in a reasonable time.

Foundational works established entanglement swapping and quantum repeater architectures to mitigate photon loss and decoherence over long distances~\cite{briegel1998quantum,dur1999quantum}. Unlike classical routing, entanglement routing must account for probabilistic link establishment, fidelity degradation through swapping and storage, and the impossibility of amplifying unknown quantum states. Consequently, routing decisions depend not only on topology and capacity, but also on the quality and temporal availability of entangled links. Van Meter \etal~\cite{Meter2021} study path selection in quantum repeater networks by adapting Dijkstra's algorithm with a link cost measured in seconds per Bell pair of a specified fidelity. Their simulations show that this cost is strongly correlated with total work, while lower-cost paths achieve higher throughput in most, but not all, tested cases. Subsequent work has explored fidelity-aware metrics and gradient-based routing rules to guide path selection under entanglement constraints~\cite{sutcliffe2025fidelity, Meter2021}. To enhance robustness and scalability, multipath and percolation-based approaches have also been explored~\cite{pirandola2019end}.

Beyond proposing new schemes to design and operate quantum networks, a number of studies formulate entanglement distribution as an optimization problem. These include multi-commodity flow-based models under capacity and fidelity constraints~\cite{Chakraborty2020}, request scheduling approaches~\cite{Cicconetti2021}, multiple-entanglement routing formulations~\cite{nguyen2022entanglement}, layered routing architectures~\cite{pant2017routing}, and concurrent flow models~\cite{shi2023concurrent}. Halder \etal~\cite{halder2024optimal}, for instance, proposed an offline routing and end-to-end entanglement distribution problem under entanglement-success, memory, and fidelity constraints. They formulate the problem as an ILP and complement it with a heuristic for larger instances, showing that the heuristic remains close to the optimum while scaling better. Nguyen \etal~\cite{nguyen2025maximizing} consider the problem of maximizing the entanglement routing rate under fidelity constraints. They first formulate the problem as an ILP, then propose approximation algorithms with theoretical guarantees, together with a more scalable practical variant. Their results show that these methods can efficiently increase the number of successfully served entanglement requests. Zeng \etal~\cite{zeng2023entanglement} address an entanglement routing design problem in a multi-user setting. They decompose the problem into two sequential integer optimization stages in order to maximize first the number of served source-destination pairs and then the expected throughput. While these approaches bring quantum routing closer to classical network optimization frameworks, most rely either on compact Integer Linear Programs (ILPs), which become computationally prohibitive at scale, or on heuristic methods offering limited guarantees on solution quality.

Recently, quantum computing has emerged as a promising paradigm for tackling combinatorial optimization problems \cite{Ciacco25, Yangyang20, Volpe25}, including shortest-path and routing-type formulations~\cite{Liu22, Osaba22}. Several approaches map shortest-path problems to QUBO models that can be addressed, for example, using quantum annealers~\cite{Krauss20} or gate-based quantum computers~\cite{Khadiev19}. These methods typically rely on sampling strategies to explore large combinatorial spaces and are often embedded within hybrid classical--quantum workflows~\cite{Coelho23}. Along the same line, hybrid quantum-classical methods have also been investigated for routing-type optimization problems. For example, Bouchmal \etal~\cite{bouchmal2024quantum} study routing in 6G optical networks through a QUBO formulation solved with QAOA in a hybrid classical--quantum loop, while Harb \etal~\cite{harb2026quantum} formulate a resilient routing problem under dual-link failures within a QAOA-based optimization framework validated on small instances. However, these works address routing in classical communication networks, not fidelity-constrained entanglement routing in quantum networks.

While such approaches have shown promise for generic combinatorial optimization and classical routing problems, their application to fidelity-constrained entanglement routing remains largely unexplored. At the same time, optimization-based entanglement routing naturally leads to large-scale path-selection problems, where the number of feasible routes may grow exponentially with the network size. This motivates decomposition techniques such as column generation, in which a restricted master problem selects routes from a limited set of candidate paths, while a pricing subproblem generates new promising routes when needed.

Compared with the above literature, our contribution does not aim to introduce a new quantum-network protocol or forwarding strategy. Rather, it targets the algorithmic path-computation bottleneck arising in the  entanglement routing optimization problem. To the best of our knowledge, this is the first work to use a neutral atom-based quantum optimization routine as a pricing oracle within a column generation framework for fidelity-constrained entanglement routing.

\section{Problem Statement and Formulations}
\label{sec:problem_statement}

In this section, we formally define the Entanglement Routing in Quantum Networks (ERQN) problem and present the mathematical formulation adopted in this work.

\subsection{Problem statement}

Let $G=(V,E)$ be an undirected graph representing the quantum network, where $V$ denotes the set of nodes, representing the network nodes (i.e., repeaters), and $E$ the set of edges, representing the communication channels within the QIN.

While every node $u \in V$ is associated with a constant fidelity decay factor $\eta\in(0,1]$, each undirected link $e\in E$ has its own capacity $C_e>0$, representing the maximum number of parallel entanglement channels, and a per-hop fidelity decay factor $\pi_e\in(0,1]$. Each node $u\in V$ is associated with a finite quantum-memory capacity $\sum_{e\in\delta(u)} C_e$, set equal to the aggregate capacity of its incident links, with $\delta(u)\subseteq E$ denoting the set of links incident to~$u$.

Let $K$ be the set of connection requests. Each request $k\in K$ is defined by a source node $s_k\in V$ and a destination node $t_k\in V$, a throughput demand $d_k>0$, representing the required number of parallel channels, and a minimum required end-to-end fidelity $F_k\in[0,1]$. Since the transmission of quantum states through optical fibers is subject to decoherence that increases exponentially with distance, resulting in an exponential decay of the end-to-end fidelity along the path~\cite{briegel1998quantum,Muralidharan2016}, each demand must be routed along a path whose cumulative fidelity remains above a minimum acceptable threshold $F_k$. In this regard, for each demand $k\in K$, let
\begin{equation*}
P_k =
\Bigl\{
p=(e_1,\dots,e_h)
\;\Big|\;
p \text{ is a simple path from } s_k \text{ to } t_k
\Bigr\}
\end{equation*}
be the set of feasible paths connecting $s_k$ to $t_k$ and respecting all requirements of demand~$k$. Also, let $h=|p|$ denote the hop count (number of edges) of~$p$. Hence, for any path $p\in P_k$, the resulting end-to-end fidelity is
\begin{equation}
\label{eq:path_fidelity}
\mathrm{Fid}(p)
=
\left(\prod_{e\in p}\pi_e\right)
\eta^{|p|-1},
\end{equation}
where the exponent $|p|-1$ accounts for the number of intermediate nodes (repeaters) performing entanglement swaps. The factor $\pi_e$ captures the fidelity degradation induced by transmission over link~$e$ (e.g., photon loss and decoherence along the corresponding physical channel), while $\eta$ captures the additional fidelity loss incurred at every intermediate repeater performing an entanglement-swapping operation. This multiplicative structure follows the standard Werner-state modeling of entangled links~\cite{Werner1989}, together with their composition under entanglement swapping~\cite{briegel1998quantum}. For simplicity, we treat $\eta$ as a single, network-wide parameter, i.e., we assume the swapping-induced degradation to be homogeneous across repeaters rather than modeling it on a per-node basis.

The Entanglement Routing in Quantum Networks (ERQN) problem can be defined as follows. Given a physical network represented by an undirected graph $G$ and a set of demands $K$, the problem aims at selecting, for each demand $k\in K$, at most one routing path so as to maximize the total number of admitted demands, subject to the following constraints:
\begin{itemize}
    \item Capacity constraints: for every undirected link $e=\{u,v\}\in E$, the total number of admitted demands routed through~$e$ does not exceed its capacity~$C_e$.
    
    \item Fidelity constraints: for every admitted request~$k$ routed along path~$p$,
    \begin{equation}
    \label{eq:werner_constraint}
    \mathrm{Fid}(p)
    \geq
    \frac{4F_k-1}{3}.
    \end{equation}
\end{itemize}
Here, $(4F_k-1)/3$ corresponds to the Werner-state fidelity threshold~\cite{Werner1989,Chakraborty2020}.

The formulation above, and the whole solution framework built upon it, is deterministic: for a given generation round, the set of available links, their capacities $C_e$, and the fidelity factors $\pi_e$ and $\eta$ are treated as known and fixed inputs. In practice, however, elementary entanglement generation and entanglement swapping both succeed with some probability, entangled pairs held in memory have a finite lifetime, and a request may need to wait for a subsequent generation round before a feasible path becomes available. This work does not model these effects explicitly, but instead targets the per-round routing decision, i.e., the assignment of requests to paths given the currently available (already heralded) links, which allows us to isolate and study the combinatorial routing/pricing bottleneck.

It is worth mentioning that, when the fidelity constraints are not considered, the ERQN problem reduces to UMCF, a known NP-hard problem~\cite{even1976complexity}, whose objective is to maximize the total number of accepted requests. Therefore, the ERQN problem is also NP-hard.

Two types of formulations have been considered in the literature to address UMCF-type problems~\cite{salimifard2022multicommodity}: the \textit{node-arc formulation} and the \textit{path formulation}. The former is a compact formulation presenting a polynomial number of variables and constraints, but it often does not scale well and may fail to handle large instances in practice using state-of-the-art dedicated solvers. The path model, on the other hand, involves an exponential number of path-related variables but can be effectively tackled with sophisticated decomposition techniques such as the column generation approach. For this reason, in what follows, we present a path-based formulation for the ERQN problem and introduce a hybrid classical-quantum algorithm to solve it.

\subsection{Path-Based Formulation}

Let us first introduce the binary decision variables
\[
x_p^k =
\begin{cases}
1, & \text{if request }k\text{ is routed}\\
   & \text{along path }p\in P_k,\\
0, & \text{otherwise}.
\end{cases}
\]

The ERQN path-based formulation (ERQN-P) is then the following:
\begin{align}
\max_{x}\quad
& \sum_{k\in K}\sum_{p\in P_k}x_p^k
\label{obj:path}\\
\text{s.t.}\quad
& \sum_{k\in K}
  \sum_{\substack{p\in P_k\\e\in p}}
  d_k x_p^k
  \le C_e,
\notag\\[-1mm]
& \hspace{25mm}\forall e\in E,
\label{cons:cap_path}\\
& \sum_{p\in P_k}
  \left[
  \sum_{e\in p}(-\ln\pi_e)
  -( |p|-1 )\ln\eta
  \right]x_p^k
\notag\\
& \quad\le
  -\ln\left(\frac{4F_k-1}{3}\right),
  \quad \forall k\in K,
\label{cons:fid_path}\\
& \sum_{p\in P_k}x_p^k\le 1,
  \quad \forall k\in K,
\label{cons:onepath}\\
& x_p^k\in\{0,1\},
  \quad \forall k\in K,\quad p\in P_k.
\label{cons:binary_path}
\end{align}

The objective~\eqref{obj:path} maximizes the total number of admitted requests. Constraint~\eqref{cons:cap_path} enforces the capacity of each undirected link by ensuring that the aggregate demand of all paths traversing the link does not exceed~$C_e$. Constraint~\eqref{cons:fid_path} corresponds to the logarithmic reformulation of the fidelity constraint~\eqref{eq:werner_constraint}, which makes the end-to-end fidelity condition additive over the links of a path. Additionally, constraint~\eqref{cons:onepath} restricts each request to at most one routing path, thereby enforcing unsplittable routing, and constraint~\eqref{cons:binary_path} defines the binary domain of variables~$x$.

Because the number of feasible paths may grow exponentially with the size of the network, solving the path-based formulation by explicit enumeration is not practical. This motivates the use of a column generation framework, presented in the next section, in which promising paths are generated dynamically, and the pricing step is assisted by a neutral-atom quantum routine.

\section{Hybrid Classical--Quantum Column Generation Method}
\label{sec:hybrid_cg}

This section describes the proposed hybrid classical--quantum column-generation approach for solving ERQN-P. The method relies on a classical column-generation scheme to control the global optimization process, while quantum resources are used to assist the pricing phase responsible for generating promising routing paths. We first present the overall framework and the restricted master problem formulation, then describe the initialization procedure, the solution algorithm of the Linear Programming (LP) relaxation, and finally the pricing mechanism together with its quantum algorithm design.

\subsection{Overall Framework}

The path-based formulation ERQN-P~\eqref{obj:path}--\eqref{cons:binary_path} involves an exponential number of variables, one for each $s_k$--$t_k$ path, and it becomes computationally expensive as the size of the instance increases. Column generation provides an efficient way to overcome this difficulty by iteratively considering only a limited subset of path variables. To this end, for each request $k\in K$, we define the set
\begin{equation}
\Phi_k
=
\left\{
p\in P_k
\;\middle|\;
\begin{array}{c}
p \text{ satisfies~\eqref{eq:werner_constraint}}\\
\text{for request }k
\end{array}
\right\},
\end{equation}
which contains all $s_k$--$t_k$ paths whose end-to-end fidelity satisfies the Werner-state threshold. Restricting formulation~\eqref{obj:path}--\eqref{cons:binary_path} to paths in~$\Phi_k$ allows the fidelity constraint to be enforced implicitly, without including constraint~\eqref{cons:fid_path} explicitly in the master problem. In a column-generation scheme, the restricted master problem (RMP) is obtained by considering only a subset of variables~$x_p^k$ corresponding to a limited number of paths in~$\Phi_k$. In this context, each path variable $x_p^k$ is viewed as a column of the master problem. This subset may initially contain only a few columns and is progressively enriched during the column-generation process. Replacing $P_k$ by $\Phi_k$ in \eqref{obj:path}--\eqref{cons:binary_path} yields the following master formulation:

\begin{align}
\max_{x}\quad
& \sum_{k\in K}\sum_{p\in\Phi_k}x_p^k
\label{obj:cg}\\
\text{s.t.}\quad
& \sum_{k\in K}
  \sum_{\substack{p\in\Phi_k\\e\in p}}
  d_k x_p^k
  \le C_e,
\notag\\[-1mm]
& \hspace{25mm}\forall e\in E,
\label{cons:cg-cap}\\
& \sum_{p\in\Phi_k}x_p^k\le 1,
  \quad \forall k\in K,
\label{cons:cg-path}\\
& x_p^k\in\{0,1\},
  \quad \forall k\in K,\quad p\in\Phi_k.
\label{cons:cg-binary}
\end{align}

We first consider the linear relaxation of this problem by replacing the binary constraints with non-negativity constraints, namely $x_p^k \ge 0$, which defines the LP relaxation denoted by ERQN-P-R. Starting from an initial RMP, the LP is solved iteratively and enriched with new path variables (columns) whenever additional paths are found that can improve the current solution. Such paths are identified through their reduced cost, which indicates whether adding a new path variable can improve the objective value while accounting for the resources it consumes. The search for these improving paths is carried out by solving the so-called pricing problem. This process continues until no path with a positive reduced cost can be found, meaning that no additional column can further improve the LP objective value. At this stage, the current solution is optimal for ERQN-P-R.

The optimal value of ERQN-P-R provides an upper bound (UB) on the original integer problem ERQN-P. Once the column-generation phase terminates, the generated paths define a restricted but high-quality search space. The integrality of the solution is then enforced by solving an integer master problem restricted to these columns, yielding a feasible integer solution for the original ERQN-P formulation.

The proposed solution framework follows the column-generation paradigm within a hybrid classical--quantum optimization architecture. Classical computation is used to solve the RMP, derive the dual values associated with its constraints, and manage the pool of generated columns, while the pricing phase, responsible for identifying promising new paths, is handled through a quantum-assisted optimization procedure. Based on these dual values, the reduced cost of each candidate path variable can be evaluated, thereby indicating whether adding the corresponding column is likely to improve the current LP solution after accounting for the resources it consumes. The overall workflow, therefore, alternates between solving the RMP, solving the pricing problem to identify improving paths, and enriching the master problem until convergence.

In practice, the method starts by generating an initial set of feasible paths using a fast heuristic in order to construct the initial RMP. The restricted master LP is then solved, and the corresponding dual values are extracted to guide the pricing phase. The quantum-assisted pricing procedure acts as a heuristic pricing oracle that searches for improving paths, which are added to the master problem whenever they have a positive reduced cost. Since failure of the quantum routine to identify an improving path does not certify that none exists, an exact classical ILP pricing phase is invoked whenever the quantum routine returns no additional column. Any improving paths found by the exact pricer are added to the RMP, and exact pricing continues until it certifies that no path with positive reduced cost remains. Once column generation has been certified in this way, a restricted integer program is finally solved over the generated columns to obtain a feasible routing solution.

\begin{figure}[t]
\centering
\includegraphics[width=\linewidth]{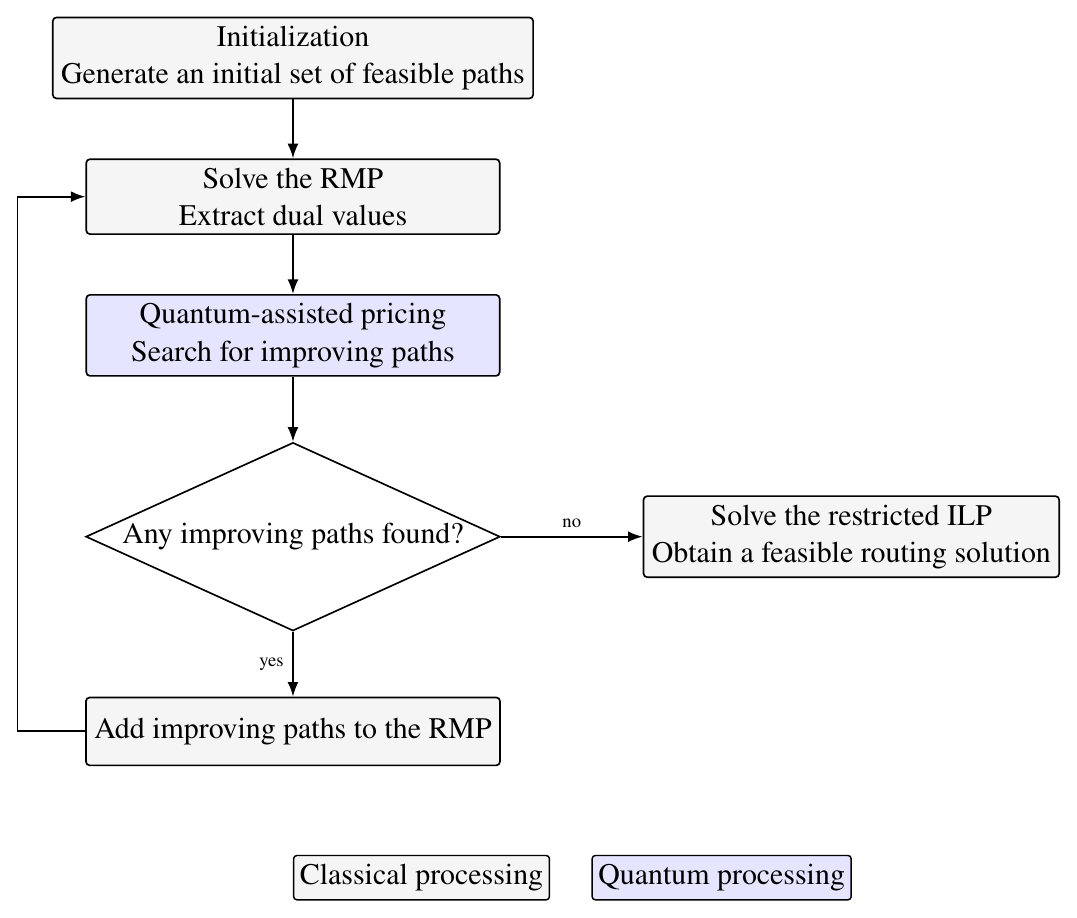}
\caption{Overall workflow of the proposed hybrid classical--quantum column-generation method. Classical computation is used for solving the RMP and the final restricted ILP, while the pricing phase is handled through a quantum-assisted optimization routine.}
\label{fig:flowchart}
\end{figure}

Figure~\ref{fig:flowchart} illustrates the overall workflow of the proposed hybrid classical--quantum column-generation method and highlights the interaction between the classical master problem and the quantum-assisted pricing phase.

The main components of the proposed hybrid column-generation framework are presented in the following subsections, covering the initialization phase, the LP-relaxation procedure, the pricing subproblem, and its quantum algorithm design.

\subsection{Initialization with a Heuristic}
\label{sec:initial_columns}

Although column generation can in principle start from an empty set of columns, in this work, the initial restricted master problem, denoted by $\text{ERQN-P-R}^{0}$, is constructed from a small set of initial path variables generated by a fast constructive heuristic. More precisely, for each request $k\in K$, the heuristic builds an initial subset $\Phi_k^{0}\subseteq\Phi_k$. The purpose of this initialization phase is to provide a meaningful starting point for the restricted master problem while keeping the number of initial variables small.

The heuristic attempts to construct feasible paths that satisfy both the link-capacity constraints and the fidelity requirement. Requests are processed sequentially, and the available capacities are updated after each accepted path. For each request $k\in K$, edges whose available capacity is smaller than $d_k$ are first removed from the working graph. A Dijkstra-type search guided by link-fidelity information is then applied to identify a candidate $s_k$--$t_k$ path. If the resulting path satisfies the fidelity requirement of request~$k$, it is added to the initial column set, and the corresponding amount of capacity is reserved along its edges. The paths obtained in this way define the initial set of columns of the restricted master problem.

The heuristic does not necessarily produce a feasible path for every request. If no path satisfying both the capacity and fidelity requirements can be found for some request $k$, then $\Phi_k^{0}=\varnothing$, and no initial column is introduced for that request. Such requests may nevertheless be considered later in the column-generation process if the pricing procedure identifies improving feasible paths.

Algorithm~\ref{alg:rene-heuristic} summarizes the initialization procedure used to generate the initial path sets~$\Phi_k^{0}$.

\par\medskip
\noindent\begin{minipage}{\columnwidth}
\refstepcounter{algorithmnumber}
\label{alg:rene-heuristic}
\textbf{Algorithm \thealgorithmnumber. Heuristic for generating initial feasible path sets $\Phi_k^{0}$}
\par\smallskip
\begin{algorithmic}[1]
\REQUIRE Graph $G=(V,E)$ with link capacities $C_e$ and link fidelity factors $\pi_e$; set of requests $K$, where each request $k\in K$ is defined by $(s_k,t_k,d_k,F_k)$; node fidelity factor $\eta$
\ENSURE Initial feasible path sets $\Phi_k^{0}$ for all $k\in K$
\FORALL{$k\in K$}
\STATE Build a working graph by removing all edges whose available capacity is smaller than $d_k$
\STATE Apply a Dijkstra-type search guided by link-fidelity information to compute a candidate $s_k$--$t_k$ path $p^{*}$
\IF{$p^{*}$ exists and satisfies~\eqref{eq:werner_constraint} for request $k$}
\STATE $\Phi_k^{0} \gets \{p^{*}\}$
\STATE Reserve $d_k$ units of capacity on each edge of $p^{*}$
\ELSE
\STATE $\Phi_k^{0} \gets \varnothing$
\ENDIF
\ENDFOR
\RETURN $\{\Phi_k^{0}\}_{k\in K}$
\end{algorithmic}
\end{minipage}
\par\medskip

\subsection{Solving the Linear Relaxation}

Once the initial RMP has been constructed using the heuristic, the column-generation procedure proceeds by solving its LP relaxation and iteratively enriching it with improving path variables.

Although Linear Programs (LPs) can practically be solved in polynomial time in their input size~\cite{khachiyan1979polynomial}, ERQN-P-R has exponentially many path variables. Column generation makes it possible to reach an optimal LP solution by iteratively adding only a promising subset of columns.

At iteration~$i$, let $\text{ERQN-P-R}^{i}$ be the RMP defined on the current column sets $\Phi_{k}^{i}\subseteq\Phi_{k}, \forall k \in K$. A solution $x_{p}^{i,k}$ of $\text{ERQN-P-R}^{i}$ induces a feasible solution of the full relaxation by setting $x_{p}^{k}=x_{p}^{i,k}$ for $p\in\Phi_{k}^{i}$ and $x_{p}^{k}=0$ otherwise.

The pricing problem, derived from LP duality~\cite{schrijver2003combinatorial}, searches for paths $p\notin\Phi_{k}^{i}$ with positive reduced cost under the current dual solution. If no such path exists, then the current solution is optimal for ERQN-P-R; otherwise, improving columns are added, and the process is repeated.

After convergence of the column-generation procedure, integrality is enforced by solving a reduced ILP over the set of generated columns. This yields a feasible integer solution for the original ERQN-P formulation.

We now describe the mathematical structure of the RMP and its associated dual formulation, which provides the dual information required to guide column generation.

\subsubsection{Restricted master and its dual}

At iteration~$i$, the restricted master LP is
\begin{align}
\max_{x}\quad
& \sum_{k\in K}
  \sum_{p\in\Phi_k^{\,i}}x_p^k
\label{eq:rm-obj}\\
\text{s.t.}\quad
& \sum_{k\in K}
  \sum_{\substack{p\in\Phi_k^{\,i}\\e\in p}}
  d_k x_p^k
  \le C_e,
\notag\\[-1mm]
& \hspace{25mm}\forall e\in E,
\label{eq:rm-cap}\\
& \sum_{p\in\Phi_k^{\,i}}x_p^k\le 1,
  \quad \forall k\in K,
\label{eq:rm-one}\\
& x_p^k\ge 0,
  \quad \forall k\in K,\quad
  p\in\Phi_k^{\,i}.
\label{eq:rm-nonneg}
\end{align}

Let $\alpha_e \ge 0$ be the dual variable associated with constraint~\eqref{eq:rm-cap} for each $e\in E$, and let $\omega_k \ge 0$ be the dual variable associated with constraint~\eqref{eq:rm-one} for each $k\in K$. The dual of \eqref{eq:rm-obj}--\eqref{eq:rm-nonneg} is
\begin{align}
\min_{\alpha,\omega}\quad
& \sum_{e\in E}C_e\alpha_e
  +\sum_{k\in K}\omega_k
\label{eq:dual-obj}\\
\text{s.t.}\quad
& \sum_{e\in p}d_k\alpha_e+\omega_k
  \ge 1,
\notag\\[-1mm]
& \hspace{7mm}\forall k\in K,\quad
  \forall p\in\Phi_k^{\,i},
\label{eq:dual-cons}\\
& \alpha_e\ge 0,
  \quad \forall e\in E,
\notag\\
& \omega_k\ge 0,
  \quad \forall k\in K.
\label{eq:dual-nonneg}
\end{align}

The optimal dual solution of the RMP provides the information needed to determine whether additional path variables can improve the current restricted solution through their reduced cost.

\subsubsection{Reduced costs and pricing condition}

Let $(\alpha^*,\omega^*)$ be an optimal solution of the dual problem \eqref{eq:dual-obj}--\eqref{eq:dual-nonneg} associated with the current RMP. Using these dual values, the reduced cost of the path variable associated with request $k$ and path $p$ is given by
\begin{align}
\bar{c}_{kp}
&=
1-
\left(
\sum_{e\in p}d_k\alpha_e^*
+\omega_k^*
\right)
\notag\\
&=
1-\omega_k^*
-d_k\sum_{e\in p}\alpha_e^*.
\label{eq:red-cost}
\end{align}

If there exists a path $p$ for some request $k$ such that $\bar{c}_{kp}>0$, then the corresponding column can improve the current restricted solution and should be added to the RMP. If no path with a positive reduced cost exists for any request, then the current restricted solution is optimal for the full relaxation. This reduced-cost characterization naturally defines the pricing problem, which consists of identifying paths with positive reduced cost. The next subsection details how this problem is formulated and solved using quantum technology.

\subsection{The Pricing Sub-Problem}
\label{sec:pricing}

At each column-generation iteration, the pricing subproblem determines whether the current RMP can be improved by adding new path variables. Given an optimal dual solution $(\alpha^*,\omega^*)$ of the current RMP, pricing searches for each request $k\in K$, for $s_k$--$t_k$ paths that satisfy the fidelity requirement and have positive reduced cost. Using the reduced-cost structure derived previously, this problem can be reformulated as a constrained shortest-path problem on a bidirected graph~\cite{festa2015constrained}. This reformulation enables an efficient search for improving columns by exploiting shortest-path techniques under fidelity constraints. If no improving path exists for any request, then the current RMP solution is optimal for the full LP relaxation. Otherwise, the corresponding columns are added to the RMP, and the procedure continues. In the proposed hybrid classical--quantum framework, the pricing phase is handled by a quantum-assisted optimization routine based on neutral-atom computing.

\subsubsection{Constrained shortest-path formulation of pricing}

Let us first introduce $H=(V,A)$, the bidirected version of graph $G=(V,E)$, where each undirected edge $\{u,v\}\in E$ induces two directed arcs $(u,v)$ and $(v,u)$ in~$A$. We assume that link capacities and fidelities are symmetric, i.e., $C_{u,v}=C_{v,u}=C_{\{u,v\}}$ and $\pi_{u,v}=\pi_{v,u}=\pi_{\{u,v\}}$. From the reduced-cost expression~\eqref{eq:red-cost}, the contribution of a path becomes additive over the arcs of the bidirected graph $H$. Combined with the fidelity feasibility condition, this yields a constrained shortest-path formulation of the pricing problem~\cite{festa2015constrained}.

For each arc $a=(u,v)\in A$, we define the arc weight
\begin{equation}
\label{eq:pricing-weight}
w_{ka}=d_k\alpha^*_{\{u,v\}},
\end{equation}
which represents the dual cost of routing request $k$ through arc $a$. For any $s_k$--$t_k$ path $p$, we then obtain
\begin{equation}
\sum_{a\in p}w_{ka}
=
d_k\sum_{e\in p}\alpha_e^*,
\end{equation}
and the reduced cost can be rewritten as
\begin{equation}
\bar{c}_{kp}
=
1-\omega_k^*
-\sum_{a\in p}w_{ka}.
\end{equation}

The key observation is that a column is improving if and only if its reduced cost, defined in~\eqref{eq:red-cost}, is positive, i.e., if $\bar{c}_{kp}>0$, which is equivalent to requiring
\begin{equation}
\sum_{a\in p}w_{ka}<1-\omega_k^*.
\end{equation}

Let $p^*$ denote an $s_k$--$t_k$ path minimizing $\sum_{a\in p} w_{ka}$. If $p^*$ does not satisfy the above inequality, then no other feasible path can satisfy it, since every other feasible path has weight greater than or equal to that of~$p^*$. Conversely, if $p^*$ satisfies this condition, then it immediately yields an improving column. Therefore, for each request $k\in K$, pricing reduces to finding a minimum-weight $s_k$--$t_k$ path under the fidelity constraint.

In addition to this reduced-cost condition, paths must also satisfy the end-to-end Werner-state fidelity constraint introduced in~\eqref{eq:werner_constraint}. Using the logarithmic reformulation of this constraint, feasibility can be expressed as
\begin{align}
& \sum_{a\in p}\bigl(-\ln\pi_a\bigr)
-(|p|-1)\ln\eta
\notag\\
& \qquad\le
-\ln\left(\frac{4F_k-1}{3}\right).
\label{eq:pricing-fidelity}
\end{align}

Combining these observations, the pricing problem for each request $k\in K$ can be formulated as the following constrained shortest-path problem:
\begin{align}
p_k^*
\in
\arg\min_{p\in P_k}
\Biggl\{&
\sum_{a\in p}w_{ka}:
\notag\\[-1mm]
&\text{$p$ satisfies~\eqref{eq:pricing-fidelity}}
\Biggr\}.
\label{eq:pricing-csp}
\end{align}

In practice, instead of generating a single column, we allow the pricing procedure to return up to $n_p$ candidate paths satisfying the fidelity constraint in order to accelerate convergence. Within our hybrid classical--quantum framework, this multi-column generation is naturally supported by the quantum-assisted optimization routine, which enables the simultaneous exploration of multiple candidate routing configurations. This capability increases the probability of identifying several improving columns during a single pricing iteration and therefore reduces the number of column-generation iterations.

To further improve the efficiency of the pricing phase, we apply preprocessing reductions that reduce the size of the graph on which the constrained shortest-path problem is solved, thereby decreasing the computational effort of the pricing procedure without affecting feasibility or optimality.

\subsubsection{Pricing pre-processing}
\label{sec:pricing-preproc}

Before solving the pricing subproblem, a request-dependent preprocessing phase is applied in order to reduce the search space. For each request $k=(s_k,t_k,d_k,F_k)\in K$, the reductions are performed on a working copy of the bidirected graph $H=(V,A)$ and are designed to preserve all feasible paths that may yield improving columns.

\paragraph{Capacity-based reduction.}

An arc $(u,v)\in A$ can belong to a feasible path for request $k$ only if the corresponding undirected edge $\{u,v\}$ has sufficient capacity to route demand $d_k$. Therefore, all arcs satisfying $C_{\{u,v\}}<d_k$ are removed.

After this filtering, vertices and arcs that do not belong to any directed path from $s_k$ to $t_k$ are removed. If $t_k$ is no longer reachable from $s_k$, then no feasible path can exist for request $k$, and the corresponding pricing problem is declared infeasible.

\paragraph{Fidelity-based reduction.}

A second reduction exploits the fidelity constraint~\eqref{eq:pricing-fidelity}. For each arc $a=(u,v)\in A$, we define the weight
\begin{equation}
c_a=-\ln(\pi_a)-\ln(\eta).
\end{equation}

Let $T_k = \frac{4F_k-1}{3}$ be the Werner-state threshold defined in~\eqref{eq:werner_constraint}. Any feasible path must satisfy
\begin{equation}
\sum_{a\in p}c_a
\le
-\ln(T_k)-\ln(\eta).
\end{equation}

Let $d_{s_k}(u)$ denote the shortest-path distance from $s_k$ to vertex $u$ with respect to the weights $c_a$, and let $d_{t_k}(u)$ denote the shortest-path distance from $u$ to $t_k$, obtained equivalently by running Dijkstra's algorithm from $t_k$ on the reversed graph. Hence, $d_{s_k}(u)+d_{t_k}(u)$ provides a lower bound on the fidelity cost of any $s_k$--$t_k$ path passing through $u$. If
\begin{equation}
d_{s_k}(u)+d_{t_k}(u)
>
-\ln(T_k)-\ln(\eta),
\end{equation}
then no $s_k$--$t_k$ path passing through $u$ can satisfy the fidelity constraint. Such vertices, together with their incident arcs, can therefore be removed.

The pricing problem is then solved on the graph obtained after applying both reductions. We denote this reduced graph by $H_k$. These preprocessing steps are repeated independently for each request and at each column-generation iteration, since the pricing problem depends on both the request parameters and the current dual solution. In the proposed hybrid classical--quantum framework, the constrained shortest-path problem defined on $H_k$ is handled by a quantum-assisted routine based on neutral-atom computing, which generates a set of candidate $s_k$--$t_k$ paths. Among these candidates, we retain up to a fixed number $n_p$ of paths that satisfy the fidelity constraint and minimize the pricing objective~\eqref{eq:pricing-csp}.

Algorithm~\ref{alg:pricing} summarizes the overall pricing procedure.

\par\medskip
\noindent\begin{minipage}{\columnwidth}
\refstepcounter{algorithmnumber}
\label{alg:pricing}
\textbf{Algorithm \thealgorithmnumber. \textsc{PricingSubproblem}$(H,k,\alpha^*,\omega^*,n_p)$}
\par\smallskip
\begin{algorithmic}[1]
\REQUIRE Bidirected graph $H=(V,A)$, request $k=(s_k,t_k,d_k,F_k)$, dual values $\alpha^*,\omega^*$, maximum number of selected paths $n_p$
\ENSURE Set of improving paths for request $k$

\STATE Construct the reduced graph $H_k$ by applying the capacity- and fidelity-based reductions
\IF{$H_k=\emptyset$ or no $s_k$--$t_k$ connection remains in $H_k$}
\STATE \RETURN $\emptyset$
\ENDIF

\FORALL{$a=(u,v)\in A(H_k)$}
\STATE $w_{ka}\leftarrow d_k\alpha^*_{\{u,v\}}$
\ENDFOR

\STATE Generate a set of candidate $s_k$--$t_k$ paths on $H_k$ using the quantum-assisted pricing routine
\STATE Select up to $n_p$ candidate paths satisfying the fidelity constraint and minimizing~\eqref{eq:pricing-csp}
\STATE $\mathcal{P}_k^{+}\leftarrow\emptyset$

\FORALL{$p\in\mathcal{P}_k$}
\IF{$\sum_{a\in p}w_{ka}<1-\omega_k^*$}
\STATE $\mathcal{P}_k^{+}\leftarrow\mathcal{P}_k^{+}\cup\{p\}$
\ENDIF
\ENDFOR

\STATE \RETURN $\mathcal{P}_k^{+}$
\end{algorithmic}
\end{minipage}
\par\medskip

\subsection{QUBO Formulation of the Pricing Sub-Problem}
\label{sec:qubo-pricing}

For each fixed request $k\in K$, the pricing problem defined in~\eqref{eq:pricing-csp} is reformulated as a Quadratic Unconstrained Binary Optimization (QUBO) problem in order to make it compatible with quantum optimization on neutral-atom devices. The proposed encoding is defined on the bidirected graph $H=(V,A)$ and uses one binary decision variable per arc.

More precisely, for each arc $(u,v)\in A$, we introduce the binary variable $z_{u,v}\in\{0,1\}$,where $z_{u,v}=1$ if arc $(u,v)$ belongs to the candidate path, and $z_{u,v}=0$ otherwise. The QUBO construction then follows the structure of the pricing problem~\eqref{eq:pricing-csp}: the objective is encoded as a linear term, while the path and fidelity requirements are enforced through penalty terms.

\subsubsection{Objective encoding}

The objective of the pricing problem~\eqref{eq:pricing-csp} is to minimize the total pricing weight of the selected path. Using the arc weights defined in~\eqref{eq:pricing-weight}, this objective is represented in binary form as
\begin{equation}
\label{eq:qubo-obj}
\sum_{(u,v)\in A}w_{k,(u,v)}z_{u,v}.
\end{equation}

\subsubsection{Flow-conservation penalty}

A necessary condition for the selected arcs to represent an $s_k$--$t_k$ path is that flow conservation holds at every node.
This means that the difference between outgoing and incoming flow must be equal to $1$ at the source, $-1$ at the destination, and $0$ at every intermediate node.
We define
\begin{equation}
b_u =
\begin{cases}
1,  & \text{if }u=s_k,\\
-1, & \text{if }u=t_k,\\
0,  & \text{otherwise},
\end{cases}
\qquad \forall u\in V.
\end{equation}
Since QUBO models do not admit explicit constraints, this condition is enforced through the quadratic penalty
\begin{align}
P_{\mathrm{flow}}(z)
=
\sum_{u\in V}\gamma_u
\Biggl(&
\sum_{\substack{v:(u,v)\in A}}z_{u,v}
-
\sum_{\substack{w:(w,u)\in A}}z_{w,u}
\notag\\
&-b_u
\Biggr)^2.
\label{eq:qubo-flow}
\end{align}

where $\gamma_u>0$ is a penalty coefficient. This term is equal to zero when flow conservation is satisfied and becomes positive as soon as the balance is violated at some node.

Since the flow-conservation penalty alone does not necessarily guarantee a simple $s_k$--$t_k$ path, a sampled binary configuration may contain cycles, disconnected components, or redundant arcs. Each binary solution is therefore post-processed by retaining the selected arcs, pruning those that do not belong to any $s_k$--$t_k$ connection, and extracting a simple $s_k$--$t_k$ path on the resulting subgraph before reduced-cost evaluation. If no simple $s_k$--$t_k$ path can be extracted from the selected arcs, the corresponding sampled configuration is discarded.

\subsubsection{Fidelity penalty}

The fidelity constraint in~\eqref{eq:pricing-csp} is represented using the logarithmic form introduced in~\eqref{eq:pricing-fidelity}. To encode this inequality within a QUBO model, we define the fidelity-violation function
\begin{align}
g_f(z)
&=
\sum_{(u,v)\in A}
\bigl(-\ln\pi_{u,v}\bigr)z_{u,v}
\notag\\
&\quad
-\ln\eta
\left(
\sum_{(u,v)\in A}z_{u,v}-1
\right)
\notag\\
&\quad
+\ln\left(\frac{4F_k-1}{3}\right).
\label{eq:qubo-gf}
\end{align}

By construction, the fidelity condition is satisfied if and only if $g_f(z)\le 0$. To penalize violations of this inequality while preserving a quadratic binary form, we adopt a second-order approximation of an exponential penalty around $0$:
\begin{equation}
\exp(g_f(z))-1
\approx
g_f(z)+\frac{1}{2}g_f(z)^2.
\end{equation}
Since $g_f(z)$ is linear in the binary variables, this approximation yields the quadratic penalty
\begin{equation}
\label{eq:qubo-fid}
P_{\mathrm{fid}}(z)
=
\mu_f
\left(
g_f(z)+\frac{1}{2}g_f(z)^2
\right),
\end{equation}
where $\mu_f>0$ controls the importance of fidelity enforcement. This construction does not require the introduction of auxiliary slack variables and therefore preserves a compact binary encoding.

\subsubsection{QUBO pricing model}

Combining the pricing objective~\eqref{eq:qubo-obj}, the flow-conservation penalty~\eqref{eq:qubo-flow}, and the fidelity penalty~\eqref{eq:qubo-fid}, we obtain the following QUBO formulation of the pricing subproblem:
\begin{align}
Q_k(z)
&=
\sum_{(u,v)\in A}
w_{k,(u,v)}z_{u,v}
\notag\\
&\quad
+\mu_f
\left(
g_f(z)+\frac{1}{2}g_f(z)^2
\right)
\notag\\
&\quad
+\sum_{u\in V}\gamma_u
\Biggl(
\sum_{\substack{v:(u,v)\in A}}z_{u,v}
\notag\\
&\qquad
-\sum_{\substack{w:(w,u)\in A}}z_{w,u}
-b_u
\Biggr)^2.
\label{eq:pricing-qubo}
\end{align}

The first term encodes the pricing objective, the second penalizes violations of the fidelity constraint, and the third penalizes violations of flow conservation. Therefore, minimizing~\eqref{eq:pricing-qubo} reduces to searching for low-weight arc selections that approximate feasible solutions of the constrained shortest-path pricing problem~\eqref{eq:pricing-csp}.

In practice, the quantum-assisted solver may return several low-energy binary configurations during a single run. After path reconstruction and reduced-cost filtering, these configurations can yield several improved columns for the same request, which contributes to accelerating the overall column-generation process.

In what follows, we describe how this QUBO model is embedded into the neutral-atom optimization workflow and how the corresponding quantum routine is designed.
\subsection{Quantum algorithm design}
\label{sec:quantum_algorithm_design}

Once the pricing problem associated with request~$k$ has been reformulated as a QUBO, the quantum workflow consists of two successive stages: \emph{register embedding} and \emph{pulse shaping}. The first stage maps the QUBO coefficients onto a physically realizable neutral-atom register, while the second stage defines the laser sequence used to drive the system toward low-energy configurations. In our setting, these two components are applied independently for each pricing QUBO instance generated during column generation.

\subsubsection{Register embedding}
\label{sec:register_embedding}

For a given pricing QUBO, register embedding consists of assigning each binary variable to a physical atom position in the neutral-atom array so that the interaction matrix induced by inter-atomic distances approximates the target QUBO coupling matrix as closely as possible. This mapping must satisfy the geometric constraints imposed by the device, including the maximum admissible distance from the origin and the minimum inter-atomic distance required to ensure experimental feasibility on the register. In addition, the set of candidate positions may be restricted to a predefined layout (e.g., square or triangular lattices), from which the embedding selects the occupied sites. Following the greedy strategy proposed in~\cite{naghmouchi2024milp}, atoms are placed sequentially: starting from a central position, each subsequent variable is assigned to the feasible unoccupied position that minimizes the cumulative deviation between the target QUBO couplings and the interactions induced with already placed atoms. The interaction matrix is updated after each placement, and the process continues until all variables are assigned.

\subsubsection{Instance-driven pulse shaping}
\label{sec:instance_driven_pulse_shaping}

Given the embedded register, the laser pulse is constructed so that the final Hamiltonian approximates the QUBO objective while preserving sufficient quantum exploration during the evolution. The driven Rydberg Hamiltonian is controlled through the site-dependent detunings $\delta_i(t)$ and the global Rabi frequency $\Omega(t)$. The detuning, together with the interaction terms, defines the classical energy landscape, while $\Omega(t)$ induces transitions between configurations and therefore controls the exploration of the solution space.

Our objective is thus to design $\delta_i(t)$ and $\Omega(t)$ so that, at the end of the sequence ($t=T$), the Hamiltonian is dominated by the problem energy and its ground state corresponds to a low-energy solution of the QUBO. The heuristic pulse-construction procedure described below is implemented in the open-source \textit{QUBO Solver} framework~\cite{pasqal_qubo_tools}.

\paragraph{Diagonal encoding.}

At the end of the sequence, the detuning and interaction terms define the classical part of the Hamiltonian. Since the interaction graph is fixed by the embedding, we encode the diagonal part of the QUBO directly in the final detuning so that the resulting energy landscape approximates the QUBO objective.

Let $Q\in\mathbb{R}^{N\times N}$ denote the pricing QUBO matrix. The target final detunings are defined as
\begin{equation}
d_i=-\alpha Q_{ii},
\label{eq:diag_encoding_clean}
\end{equation}
where $\alpha>0$ is a scaling factor chosen so that both the detuning amplitudes and the pulse parameters remain within hardware limits. Defining $d_{\min}=\min_i d_i$, $d_{\max}=\max_i d_i$, and $\Delta_d=d_{\max}-d_{\min}$, the final detuning profile is implemented by decomposing the control into a global detuning component and a local detuning component enabled by Detuning Map Modulator (DMM)~\cite{pulser_dmm}, which provides site-dependent energy offsets of the form
\begin{equation}
\delta_i(t)
=
\delta_g(t)
+
\delta_{\mathrm{DMM}}(t)w_i.
\end{equation}
For $\Delta_d>0$, we set $\delta_g(T)=d_{\max}$, $\delta_{\mathrm{DMM}}(T)=-\Delta_d$, and
\begin{equation}
w_i=\frac{d_{\max}-d_i}{\Delta_d},
\end{equation}
which yields $\delta_i(T)=d_i$ for all $i$. In the degenerate case $\Delta_d=0$, no local correction is required and a uniform detuning is used.

\paragraph{Quantum exploration strength.}

While the detuning defines the final problem landscape, the Rabi frequency controls the exploration of the configuration space. If $\Omega(t)$ is too small, the dynamics remains close to the initial state and the exploration is limited. Conversely, if it is too large, the driving term dominates the problem Hamiltonian and the system becomes less sensitive to the energy differences that distinguish high-quality solutions.

To balance these two effects, we choose the characteristic amplitude of the pulse from the energy scale induced by the diagonal encoding:
\begin{equation}
\Omega_{\max}
=
\kappa\max_i|d_i|,
\label{eq:omega_final_clean}
\end{equation}
where $\kappa$ is a dimensionless coefficient controlling the balance between exploration and energy selectivity. In practice, $\kappa$ is chosen so that the pulse remains within hardware limits while keeping the driving term on the same order of magnitude as the classical energy scale.

\begin{figure}[t]
\centering
\includegraphics[
    width=\linewidth,
    height=0.2\textheight,
    keepaspectratio
]{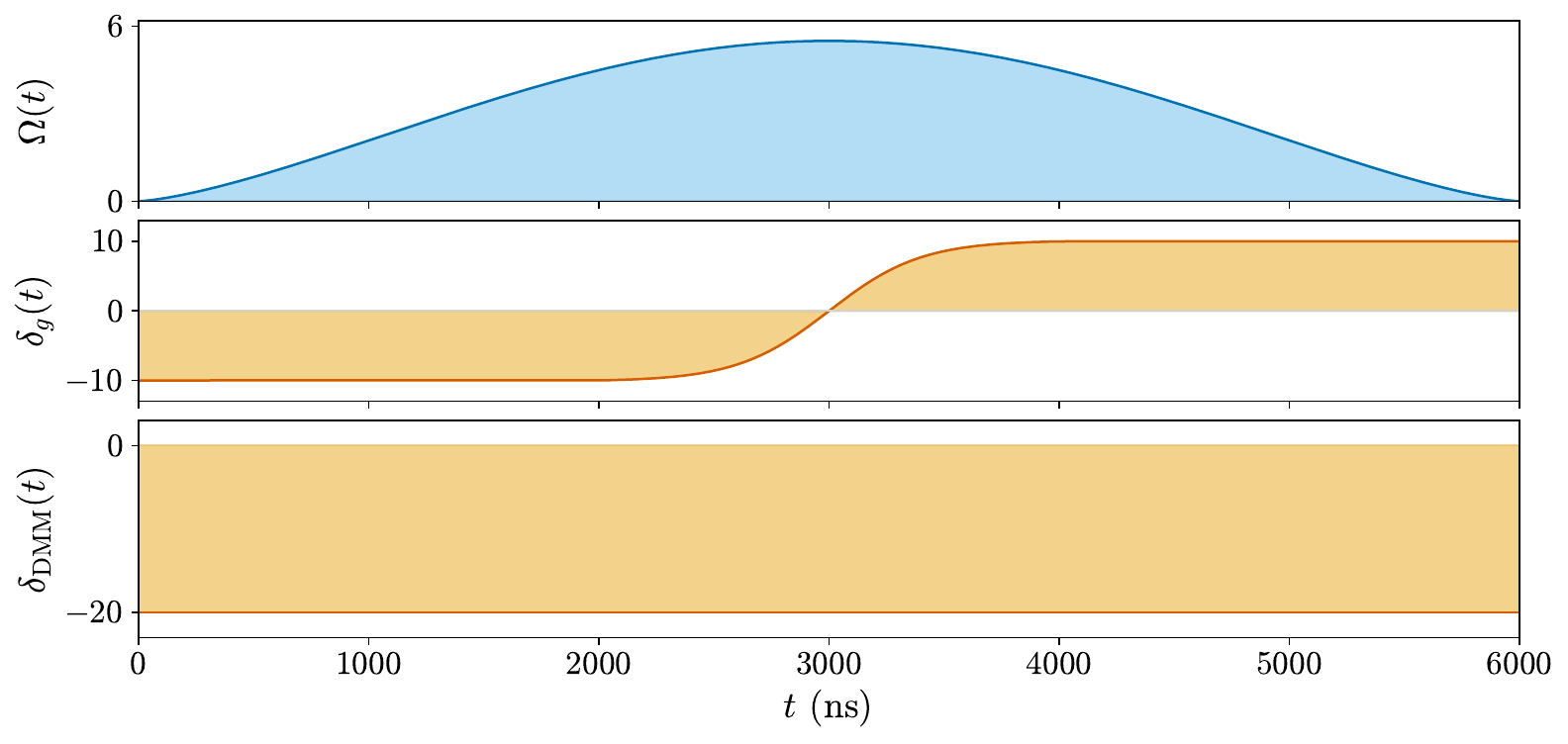}
\caption{Example of a pulse generated by the proposed instance-driven pulse shaping procedure. Top: global Rabi amplitude $\Omega(t)$ controls quantum mixing. Middle: global detuning $\delta_g(t)$ defines the classical energy landscape. Bottom: local detuning contribution $\delta_{\mathrm{DMM}}(t)$ used to encode the diagonal QUBO coefficients. The sequence starts from a strongly negative detuning to prepare the initial state and progressively transitions toward a problem-dominated Hamiltonian.}
\label{fig:pulse-example}
\end{figure}

The sequence of duration $T$ is then defined by simple four-point waveforms:
\begin{align}
\Omega(t)
&:
[\varepsilon,\Omega_{\max},
\Omega_{\max},\varepsilon],
\\
\delta_g(t)
&:
[\delta_g^{\min},\delta_g^{\min},
d_{\max},d_{\max}],
\\
\delta_{\mathrm{DMM}}(t)
&:
-\Delta_d,
\end{align}
where the arrays denote the waveform values at the four uniformly spaced time points $\{0,T/3,2T/3,T\}$. The DMM contributes a time-constant detuning $-\Delta_d$, weighted per site by $w_i$, so that atom $i$ experiences the total local detuning $\delta_i(t)=\delta_g(t)+w_i\delta_{\mathrm{DMM}}(t)$. Starting from a strongly negative detuning ensures reliable preparation of the all-zero state, while the final configuration enforces the desired diagonal encoding of the QUBO. An example of the resulting pulse schedule is shown in Fig.~\ref{fig:pulse-example}.

\section{Performance evaluation}
\label{sec:numerical_results}

This section presents the experimental protocol and the corresponding performance evaluation. We first describe the benchmark instances, the compared methods, and the evaluation metrics, and then report the results obtained by the proposed hybrid classical--quantum column-generation framework and the considered classical baselines.

\subsection{Experimental Settings}

Our experimental campaign is organized around three components: benchmark instance generation, pricing backends and compared methods, and evaluation metrics.

\paragraph{Benchmark graph instances.}

Experiments are conducted on two fixed reference topologies. Topology~1 is a sparse graph with 30 nodes and 44 undirected edges, corresponding to a density of $\rho=0.10$. Topology~2 is denser, with 12 nodes and 23 undirected edges, corresponding to a density of $\rho=0.34$.

From each reference topology, we generate connected subgraphs of increasing size $N$ at five normalized density levels,
\begin{equation}
\rho_{\mathrm{norm}}
\in
\{0.2,0.4,0.6,0.8,1.0\}.
\end{equation}
The case $\rho_{\mathrm{norm}}=1.0$ corresponds to the original reference topology when $N$ reaches the size of the underlying topology. This construction yields 25 graph configurations per topology, obtained as the Cartesian product of five graph sizes and five density levels. For each configuration, 20 independent instances are generated, resulting in a total of 500 instances per topology.

Each instance is evaluated under a traffic scenario composed of 50 routing demands. To avoid trivially feasible cases and to expose meaningful optimization trade-offs, link capacities and per-request channel requirements are generated so as to induce a sufficiently loaded operating regime. Each request is further assigned a stringent end-to-end fidelity requirement, with threshold $F_k>0.9$. To reflect the physical degradation of entanglement quality over distance, the link fidelity factors $\pi_e$ are generated according to an exponentially decaying law with respect to distance, consistently with the Werner-state-based fidelity model introduced in Sec.~\ref{sec:problem_statement}.

\paragraph{Pricing backends and compared methods.}

For the proposed hybrid column-generation framework, the pricing problem is reformulated as a QUBO and solved using the \textit{QUBO Solver} framework~\cite{pasqal_qubo_tools}, which implements the neutral-atom optimization workflow described in Sec.~\ref{sec:quantum_algorithm_design}. Depending on the size of the pricing instance, the underlying quantum simulation relies either on the statevector (SV) emulator~\cite{pasqal_emusv} or, for larger QUBO instances with more than 25 logical variables, on the matrix product-state (MPS) emulator~\cite{pasqal_emumps}. Before pricing is solved, the request-dependent preprocessing reductions described in Sec.~\ref{sec:pricing-preproc} are applied.

We compare the following methods.

\begin{itemize}
\item \textbf{CG-Q}: the proposed hybrid classical--quantum column-generation framework, in which the pricing QUBO is solved through the neutral-atom workflow implemented in \textit{QUBO Solver}~\cite{pasqal_qubo_tools}.

\item \textbf{CG-SA}: a classical column-generation baseline obtained by replacing the quantum pricing backend with simulated annealing, while keeping the same column-generation structure, pricing QUBO formulation, and preprocessing steps. Simulated annealing is also executed through \textit{QUBO Solver}~\cite{pasqal_qubo_tools}. This baseline is considered because it provides a stochastic classical heuristic for the same QUBO pricing problem, and therefore constitutes a direct classical counterpart to the sampling-based quantum pricing routine. In addition, simulated annealing is a widely used and well-established heuristic for QUBO optimization, which makes it a relevant classical reference in this setting.

\item \textbf{Greedy}: the constructive heuristic introduced in Sec.~\ref{sec:initial_columns}. This heuristic is used both to generate the initial column set in the warm-start configurations and as an independent classical baseline, in order to assess the benefit of the full column-generation framework relative to the initialization heuristic itself.
\end{itemize}

For both CG-Q and CG-SA, we consider four configurations: no reinforcement, warm-start (WS) only, post-processing (PP) only, and warm-start combined with post-processing. Although WS is treated here as a reinforcement mechanism, it remains intrinsically tied to the column-generation procedure since the overall performance depends on the quality of the initial column set generated by the greedy heuristic in Sec.~\ref{sec:initial_columns}. The PP configuration applies a classical refinement step to the sampler outputs after the column-generation loop.

As a reference for solution quality, we use the Compact ILP formulation given in Appendix~\ref{app:compact_ilp}, solved to optimality with CPLEX~\cite{cplex_chapter}. The corresponding optimal objective value is used in the gap computation.

\paragraph{Evaluation metrics.}

We analyze two complementary performance metrics. First, we report the mean number of column-generation iterations, which quantifies how many pricing rounds each column-generation method performs before the CG loop terminates. Since the Greedy heuristic does not rely on column generation, it is excluded from this analysis.

Second, solution quality is assessed through the optimality gap
\begin{equation}
\mathrm{Gap}(\%)
=
\frac{Z^{\star}-Z}{Z^{\star}}
\times 100,
\end{equation}
where $Z^{\star}$ denotes the objective value obtained from the Compact ILP formulation solved with CPLEX, and $Z$ is the objective value of the integer solution returned by the method under evaluation.

All reported results are averaged over the 20 instances associated with each graph configuration. Error bars, when shown, represent $95\%$ confidence intervals.

\subsection{Column-generation iterations}

\begin{figure*}[t]
\centering
\begin{subfigure}[t]{0.24\linewidth}
\centering
\includegraphics[width=\linewidth]{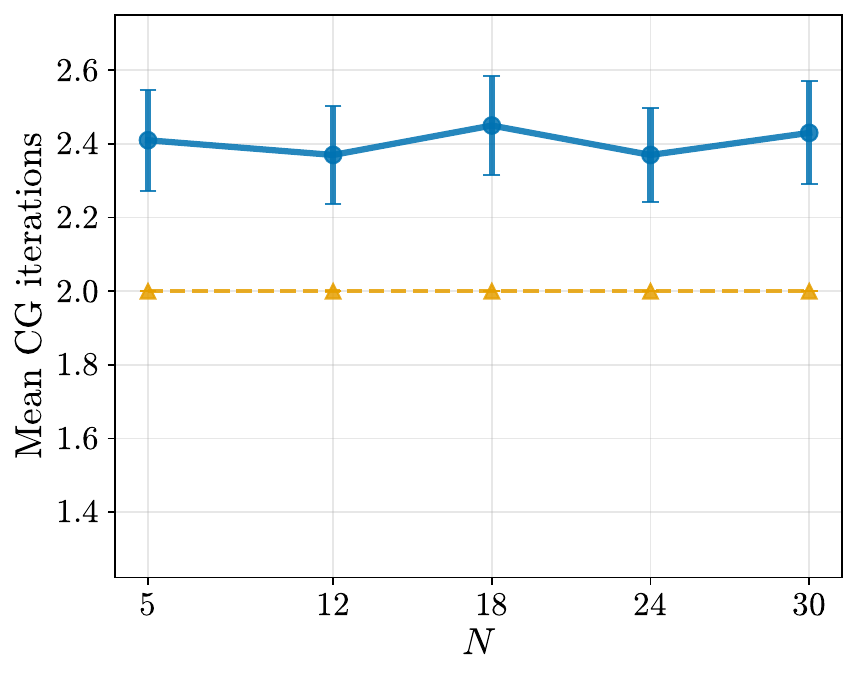}
\caption{No reinforcement.}
\label{fig:t1_50_base_it}
\end{subfigure}
\hfill
\begin{subfigure}[t]{0.24\linewidth}
\centering
\includegraphics[width=\linewidth]{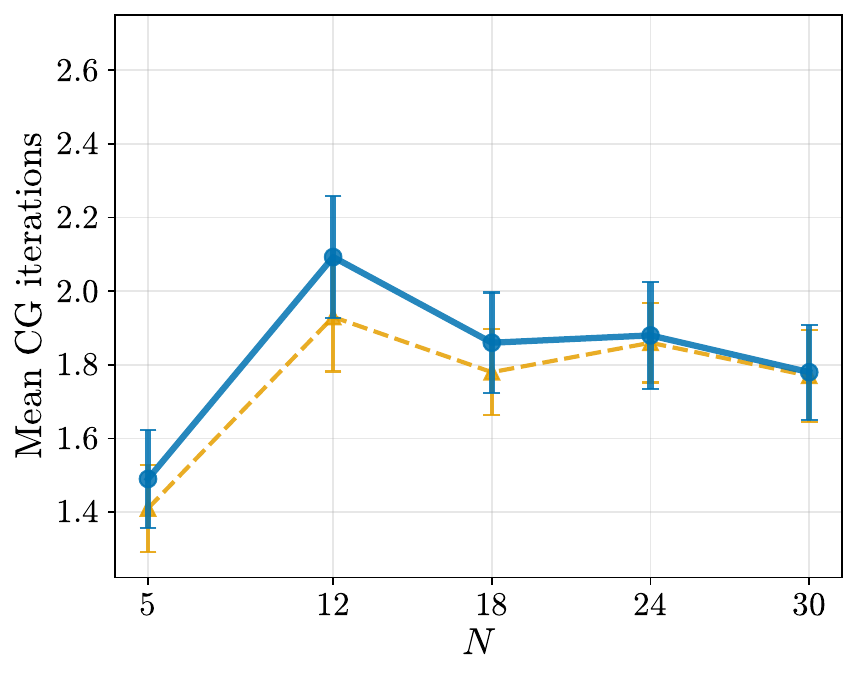}
\caption{Warm-start only.}
\label{fig:t1_50_ws_it}
\end{subfigure}
\hfill
\begin{subfigure}[t]{0.24\linewidth}
\centering
\includegraphics[width=\linewidth]{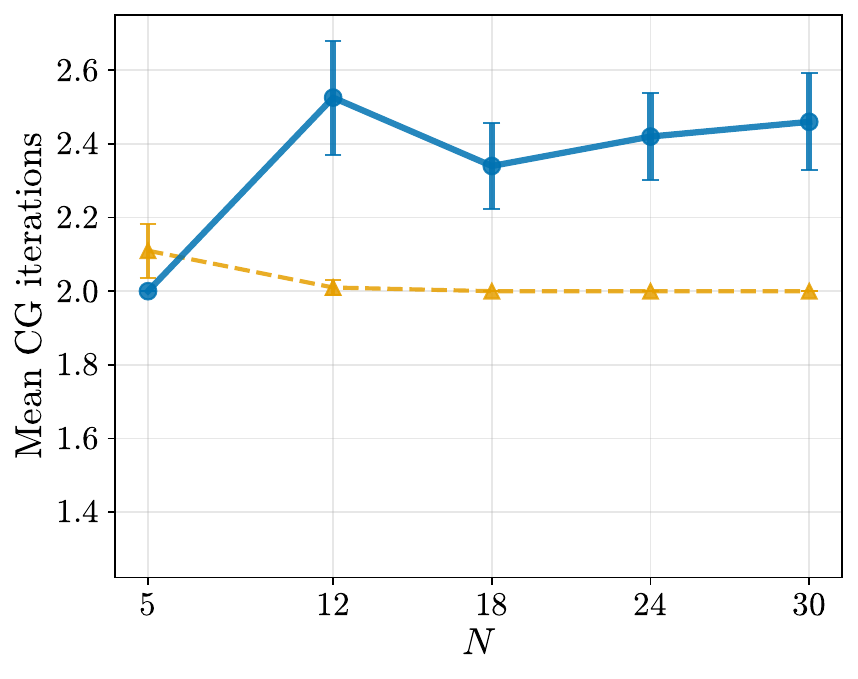}
\caption{Post-processing only.}
\label{fig:t1_50_pp_it}
\end{subfigure}
\hfill
\begin{subfigure}[t]{0.24\linewidth}
\centering
\includegraphics[width=\linewidth]{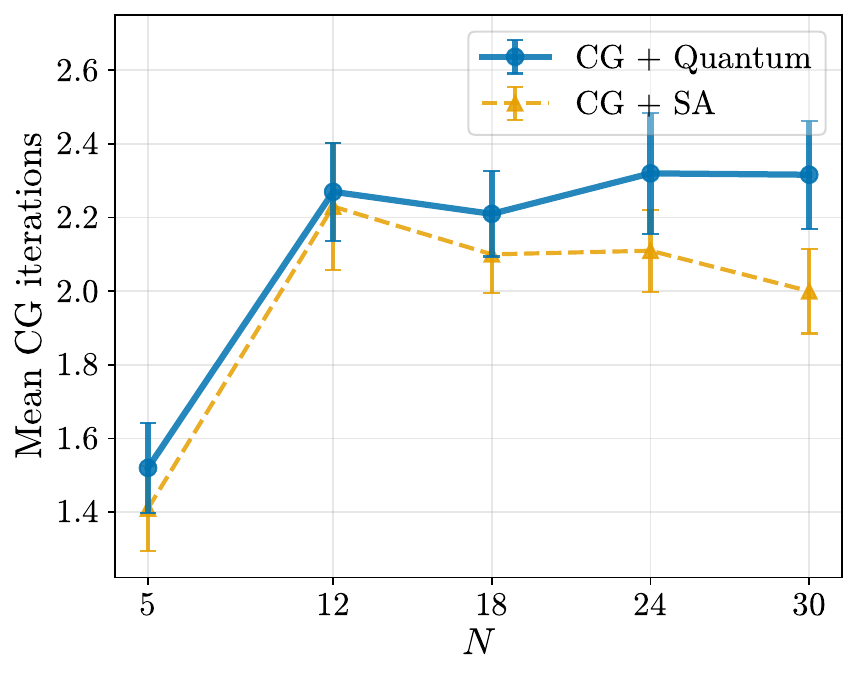}
\caption{Warm-start + post-processing.}
\label{fig:t1_50_wspp_it}
\end{subfigure}
\caption{Mean column-generation iterations vs.\ network size~$N$ on Topology~1 (20 instances per size $N$). (a)~No reinforcement: CG-SA is flat at~$2$, whereas CG-Q stabilizes around~$2.4$. (b)~Warm-start only: both methods grow together from~${\sim}1.5$ to~$2.0$--$2.1$. (c)~Post-processing only: profiles are virtually identical to the base case. (d)~Warm-start + post-processing: the trend closely matches the WS-only case.}
\label{fig:t1_50_all_it}
\end{figure*}

\begin{figure*}[t]
\centering
\begin{subfigure}[t]{0.24\linewidth}
\centering
\includegraphics[width=\linewidth]{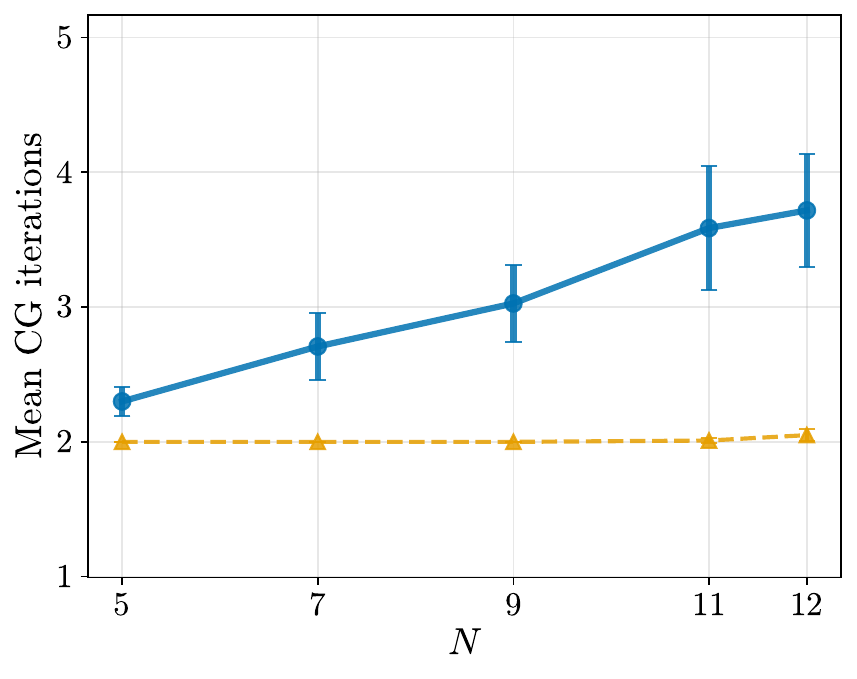}
\caption{No reinforcement.}
\label{fig:t2_50_base_it}
\end{subfigure}
\hfill
\begin{subfigure}[t]{0.24\linewidth}
\centering
\includegraphics[width=\linewidth]{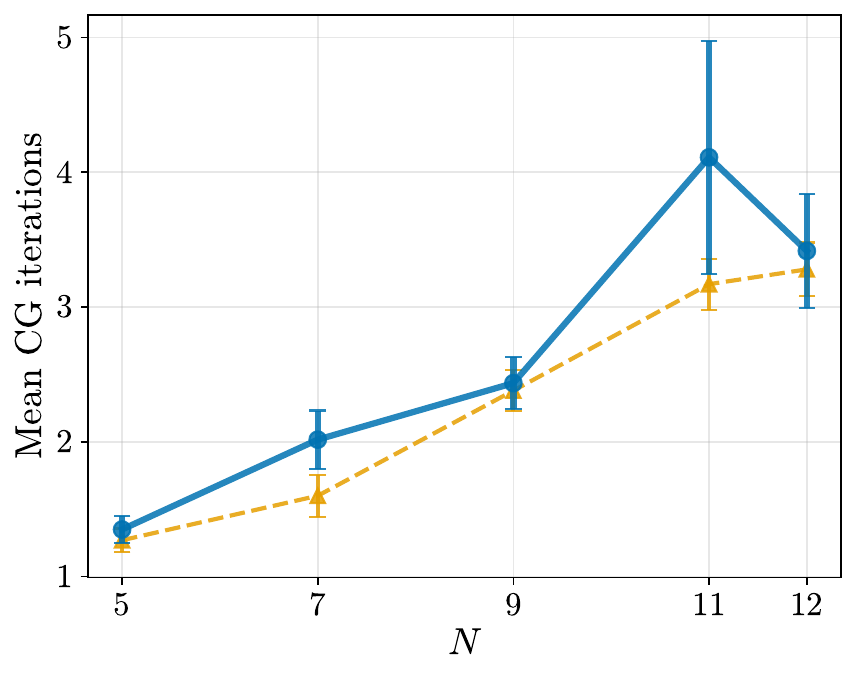}
\caption{Warm-start only.}
\label{fig:t2_50_ws_it}
\end{subfigure}
\hfill
\begin{subfigure}[t]{0.24\linewidth}
\centering
\includegraphics[width=\linewidth]{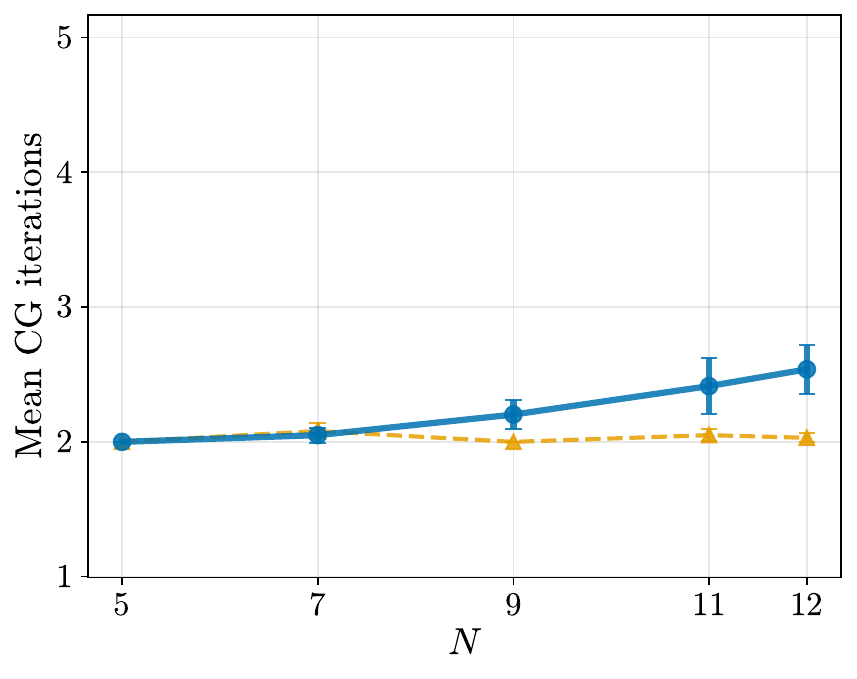}
\caption{Post-processing only.}
\label{fig:t2_50_pp_it}
\end{subfigure}
\hfill
\begin{subfigure}[t]{0.24\linewidth}
\centering
\includegraphics[width=\linewidth]{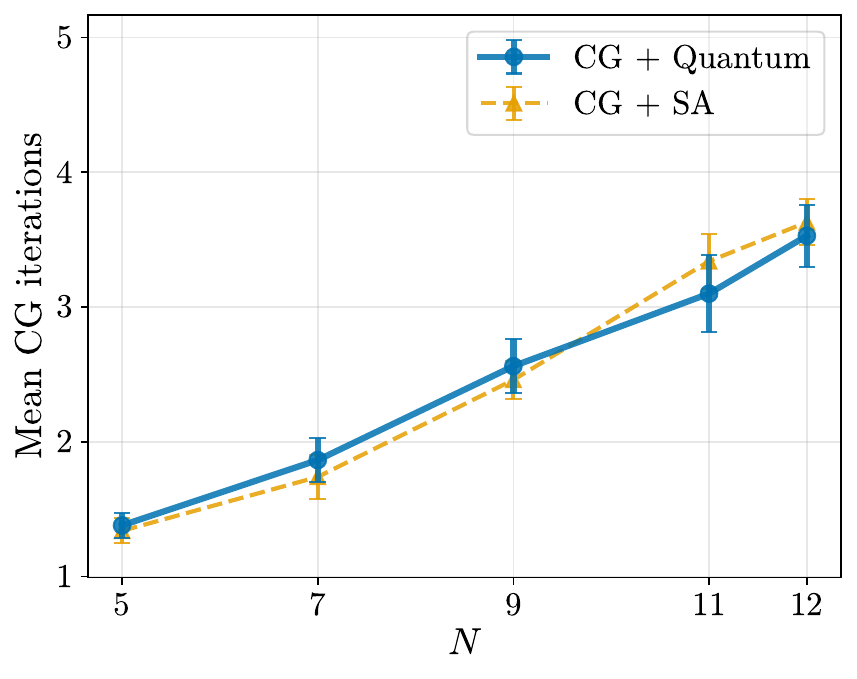}
\caption{Warm-start + post-processing.}
\label{fig:t2_50_wspp_it}
\end{subfigure}
\caption{Mean column-generation iterations vs.\ network size~$N$ on Topology~2 (20 instances per size $N$). (a)~No reinforcement: CG-SA remains at~$2$, whereas CG-Q rises to~$3.7$ at $N{=}12$. (b)~Warm-start only: both methods grow steeply to~$3.3$--$3.4$. (c)~Post-processing only: profiles remain close to the base case. (d)~Warm-start + post-processing: the trend closely matches the WS-only case.}
\label{fig:t2_50_all_it}
\end{figure*}

Figures~\ref{fig:t1_50_all_it} and~\ref{fig:t2_50_all_it} report the mean number of column-generation iterations for CG-Q and CG-SA as a function of network size~$N$.

In the base configuration, i.e., without reinforcement (Figs.~\ref{fig:t1_50_base_it} and~\ref{fig:t2_50_base_it}), CG-SA terminates in exactly~$2$ iterations for every~$N$ on both topologies, with negligible variance. This flat behavior indicates that simulated annealing fails to identify an improving column after the first sampling round, which leads to an early termination of the column-generation loop. By contrast, CG-Q sustains the column-generation process for additional rounds. On Topology~1, it stabilizes around~$2.4$ iterations, whereas on Topology~2 it rises from approximately~$2.3$ at $N{=}5$ to around~$3.7$ at $N{=}12$. This behavior shows that the quantum pricing routine continues to identify improving columns beyond those found by CG-SA. Adding post-processing alone (Figs.~\ref{fig:t1_50_pp_it} and~\ref{fig:t2_50_pp_it}) leaves these profiles nearly unchanged: CG-SA remains at~$2$, whereas CG-Q increases only slightly, to around~$2.5$ iterations.

Introducing warm-start, as described in Sec.~\ref{sec:initial_columns}, changes this behavior. With warm-start only (Figs.~\ref{fig:t1_50_ws_it} and~\ref{fig:t2_50_ws_it}), CG-SA is no longer locked at~$2$ iterations and instead grows with~$N$, closely tracking CG-Q. On Topology~1, both methods range from roughly~$1.5$ at $N{=}5$ to~$1.8$--$2.1$ at $N{=}30$, with largely overlapping confidence intervals. On Topology~2, the growth is steeper, from about~$1.3$ to~$3.3$--$3.4$ at $N{=}12$, although CG-Q exhibits a local peak around~$4.1$ at $N{=}11$, with wider uncertainty. Warm-start provides both methods with a richer initial column pool, thereby preventing the premature termination observed in the base configuration. Combining warm-start with post-processing (Figs.~\ref{fig:t1_50_wspp_it} and~\ref{fig:t2_50_wspp_it}) yields iteration profiles that remain close to the WS-only case: on Topology~1, both methods grow from about~$1.5$ to~$2.0$; on Topology~2, they increase in near-lockstep to around~$3.5$.

Overall, CG-Q requires the same or slightly more column-generation iterations than CG-SA. This behavior indicates that the quantum pricing routine continues to identify improving columns in regimes where CG-SA stops earlier. As discussed in the next subsection, this richer column pool is what ultimately enables the improved final solution quality obtained with CG-Q.

\subsection{Solution quality}

\begin{figure*}[t]
\centering
\begin{subfigure}[t]{0.24\linewidth}
\centering
\includegraphics[width=\linewidth]{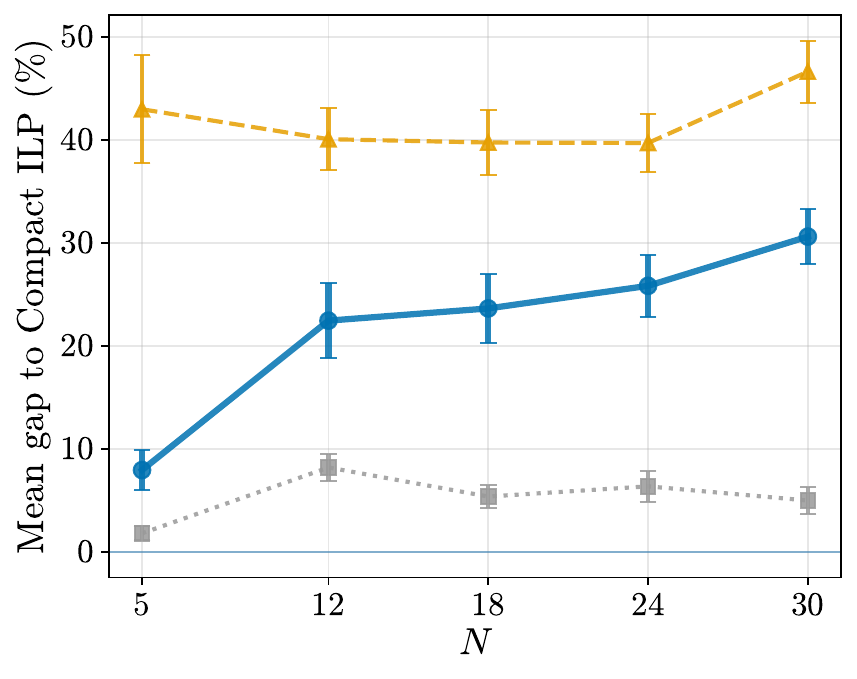}
\caption{No reinforcement.}
\label{fig:t1_50_base}
\end{subfigure}
\hfill
\begin{subfigure}[t]{0.24\linewidth}
\centering
\includegraphics[width=\linewidth]{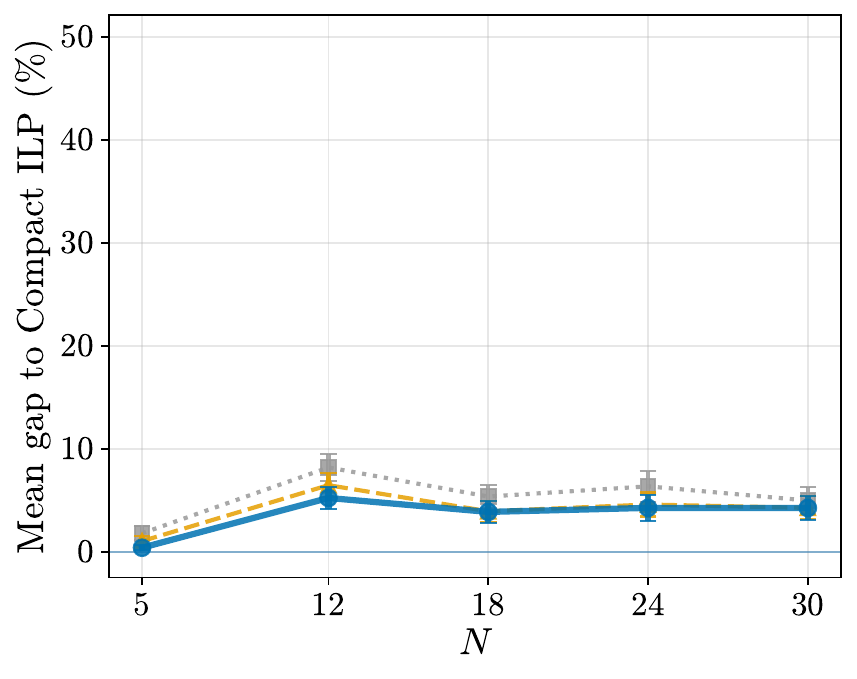}
\caption{Warm-start only.}
\label{fig:t1_50_ws}
\end{subfigure}
\hfill
\begin{subfigure}[t]{0.24\linewidth}
\centering
\includegraphics[width=\linewidth]{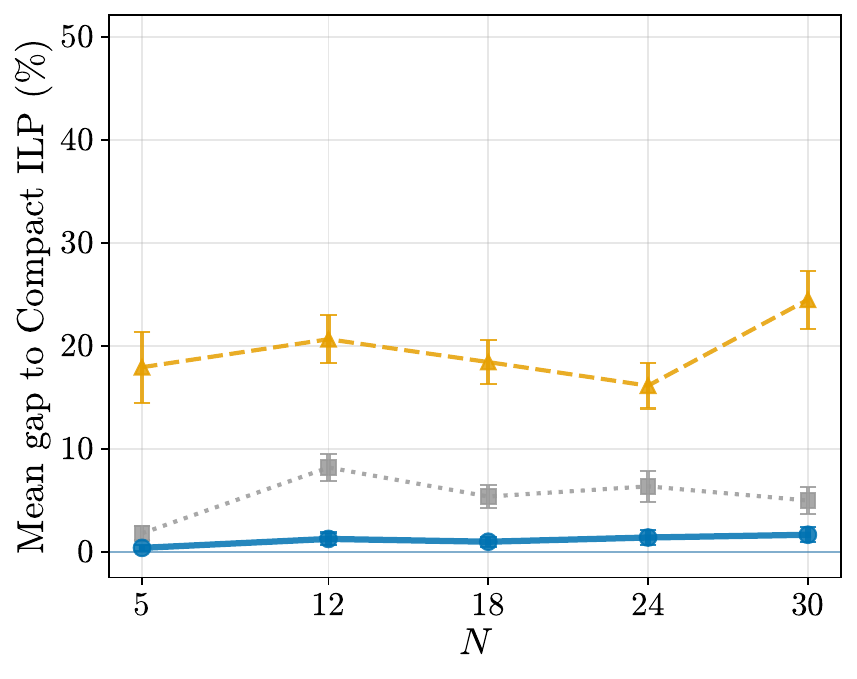}
\caption{Post-processing only.}
\label{fig:t1_50_pp}
\end{subfigure}
\hfill
\begin{subfigure}[t]{0.24\linewidth}
\centering
\includegraphics[width=\linewidth]{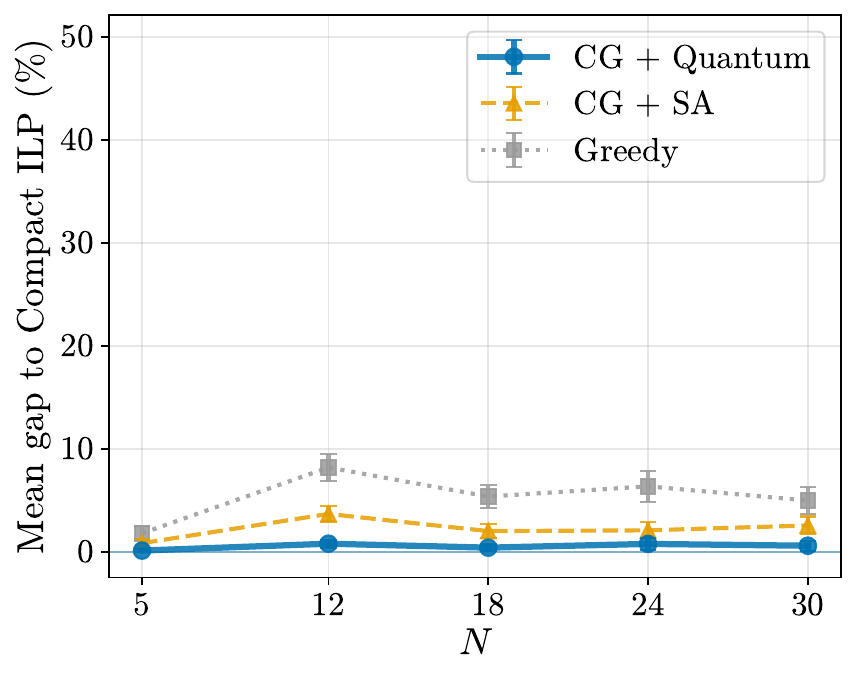}
\caption{Warm-start + post-processing.}
\label{fig:t1_50_wspp}
\end{subfigure}
\caption{Mean optimality gap to the compact ILP reference (\%) vs.\ network size~$N$ on Topology~1 (20 instances per size $N$). (a)~No reinforcement: CG-SA at~$40$--$47\%$, CG-Q at~$8$--$30\%$, Greedy at~$2$--$8\%$. (b)~Warm-start only: both CG methods converge to~$3$--$6\%$. (c)~Post-processing only: CG-Q drops below~$2\%$, whereas CG-SA remains at~$15$--$25\%$. (d)~Warm-start + post-processing: CG-Q remains below~$1\%$, whereas CG-SA is at~$1$--$3\%$.}
\label{fig:t1_50_all_gap}
\end{figure*}

\begin{figure*}[t]
\centering
\begin{subfigure}[t]{0.24\linewidth}
\centering
\includegraphics[width=\linewidth]{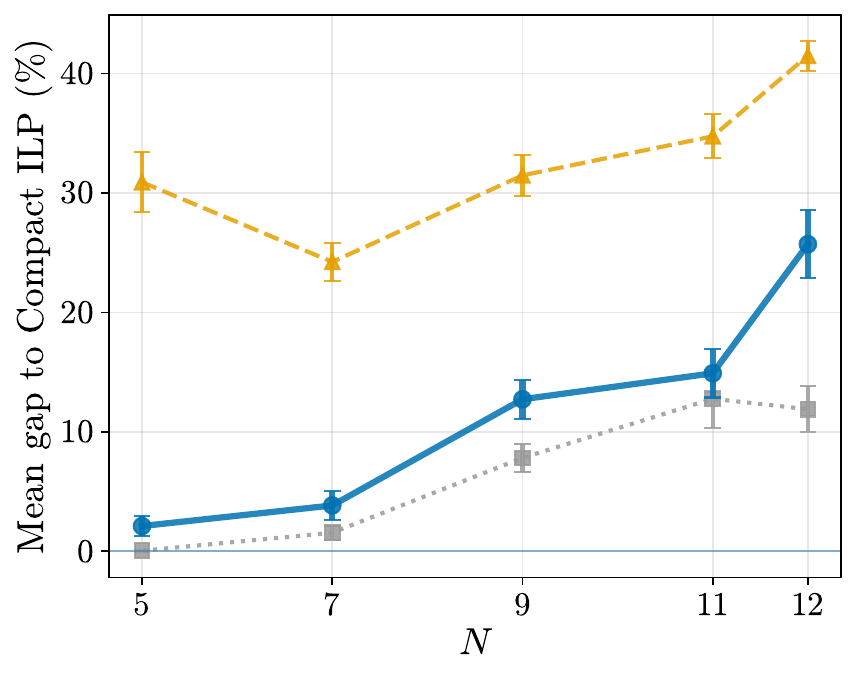}
\caption{No reinforcement.}
\label{fig:t2_50_base}
\end{subfigure}
\hfill
\begin{subfigure}[t]{0.24\linewidth}
\centering
\includegraphics[width=\linewidth]{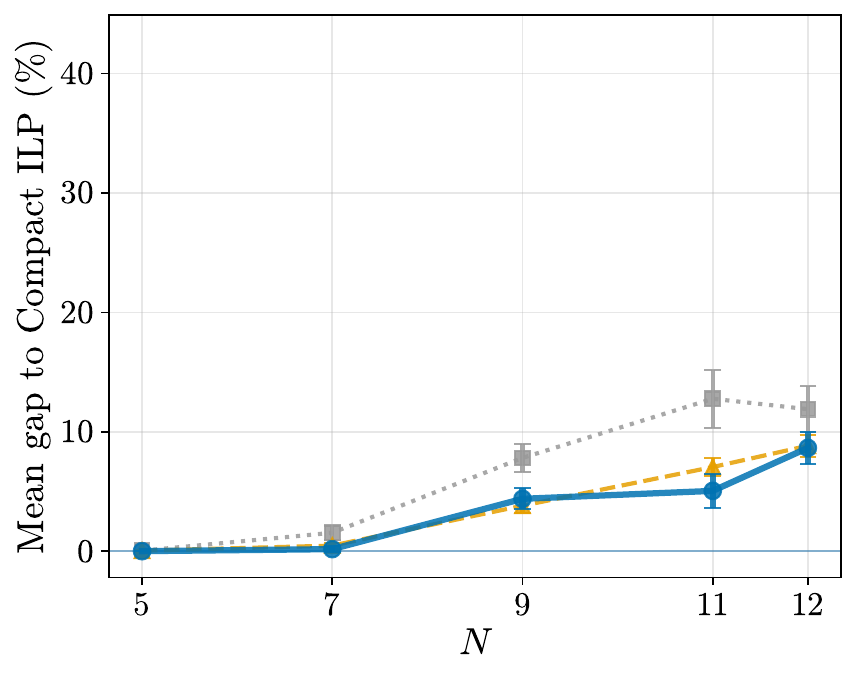}
\caption{Warm-start only.}
\label{fig:t2_50_ws}
\end{subfigure}
\hfill
\begin{subfigure}[t]{0.24\linewidth}
\centering
\includegraphics[width=\linewidth]{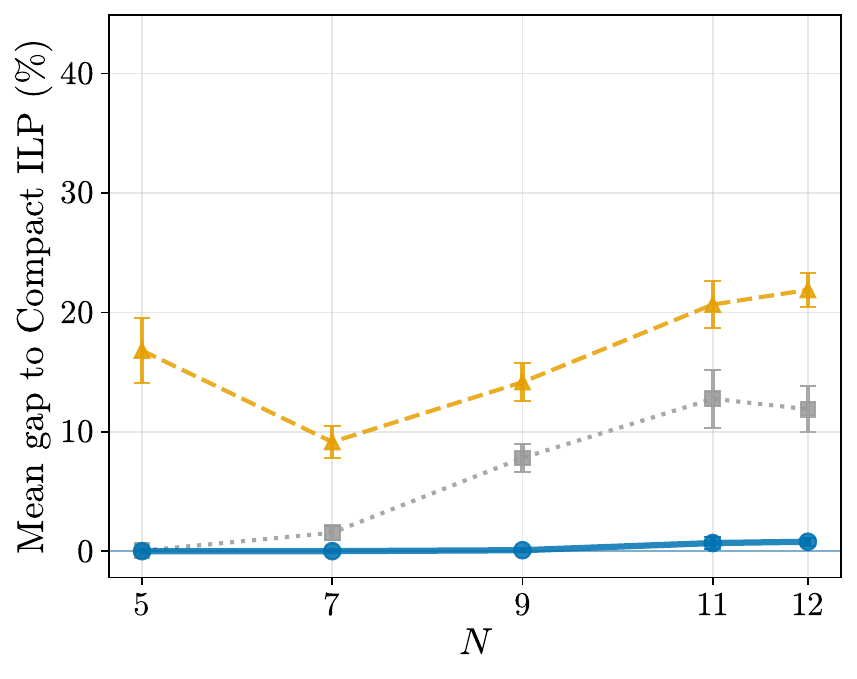}
\caption{Post-processing only.}
\label{fig:t2_50_pp}
\end{subfigure}
\hfill
\begin{subfigure}[t]{0.24\linewidth}
\centering
\includegraphics[width=\linewidth]{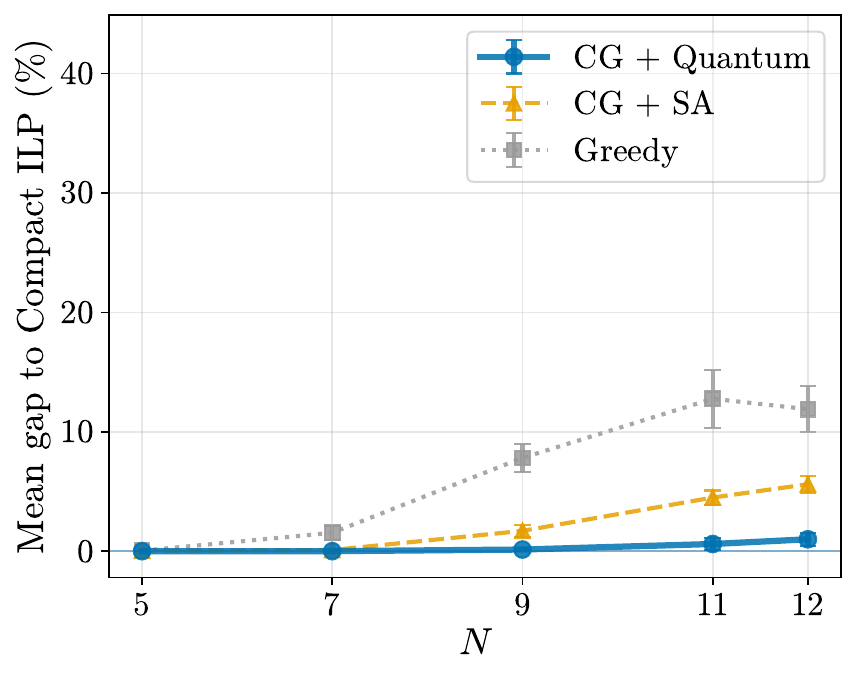}
\caption{Warm-start + post-processing.}
\label{fig:t2_50_wspp}
\end{subfigure}
\caption{Mean optimality gap to the compact ILP reference (\%) vs.\ network size~$N$ on Topology~2 (20 instances per size $N$). (a)~No reinforcement: CG-SA at~$24$--$41\%$, CG-Q at~$2$--$25\%$, Greedy at~$1$--$12\%$. (b)~Warm-start only: both CG methods remain below~$9\%$. (c)~Post-processing only: CG-Q remains below~$1\%$, whereas CG-SA is around~$22\%$. (d)~Warm-start + post-processing: CG-Q remains below~$1\%$, whereas CG-SA lies between~$0$ and~$6\%$.}
\label{fig:t2_50_all_gap}
\end{figure*}

Figures~\ref{fig:t1_50_all_gap} and~\ref{fig:t2_50_all_gap} report the mean optimality gap for all three methods with respect to the optimal value obtained by solving the Compact ILP formulation with CPLEX.

Without reinforcement (Figs.~\ref{fig:t1_50_base} and~\ref{fig:t2_50_base}), CG-SA exhibits the largest gaps, approximately~$40$--$47\%$ on Topology~1 and~$24$--$41\%$ on Topology~2, with the deficit increasing at larger~$N$. CG-Q performs better in this regime, although its gap also increases with instance size, reaching about~$30\%$ on Topology~1 at $N{=}30$ and about~$25\%$ on Topology~2 at $N{=}12$. In these base configurations, the Greedy heuristic outperforms both column-generation variants, with gaps ranging from~$2$ to~$8\%$ on Topology~1 and from~$1$ to~$12\%$ on Topology~2. This behavior is consistent with the iteration analysis above: when the column-generation process stops too early, the full benefit of the framework cannot be realized.

Adding warm-start alone (Figs.~\ref{fig:t1_50_ws} and~\ref{fig:t2_50_ws}) improves both column-generation methods. On Topology~1, CG-Q and CG-SA both fall within a narrow band of~$3$--$6\%$, comparable to Greedy. On Topology~2, both remain below~$9\%$ at $N{=}12$, whereas Greedy reaches~$12\%$. In this regime, the difference between CG-Q and CG-SA becomes small, which indicates that a stronger initialization reduces the gap between the two pricing backends.

The post-processing-only panels (Figs.~\ref{fig:t1_50_pp} and~\ref{fig:t2_50_pp}) show a different pattern. Applying post-processing to the outputs of CG-Q alone, without warm-start, is sufficient to reduce the gap to below~$2\%$ on Topology~1 and below~$1\%$ on Topology~2 over the full range of~$N$. The same refinement applied to CG-SA leaves substantially larger residual gaps, namely~$15$--$25\%$ on Topology~1 and around~$22\%$ on Topology~2, while Greedy remains between~$5$ and~$12\%$. This comparison shows that the effect of post-processing is strongly dependent on the quality of the candidate columns produced by the pricing routine. In particular, the outputs generated by CG-Q are structurally better suited to classical refinement than those produced by CG-SA.

This behavior is particularly visible on Topology~2, where post-processing alone brings CG-Q very close to the compact ILP reference, while CG-SA remains clearly above it. The combined warm-start and post-processing configuration (Figs.~\ref{fig:t1_50_wspp} and~\ref{fig:t2_50_wspp}) yields the most robust overall behavior: CG-Q remains below~$1\%$ on both topologies and across all tested sizes, whereas CG-SA reaches about~$1$--$3\%$ on Topology~1 and about~$0$--$6\%$ on Topology~2. Greedy, which is unaffected by the CG reinforcements, remains between~$5$ and~$12\%$. For example, at $N{=}30$ on Topology~1, CG-Q achieves a gap of about~$0.8\%$ versus about~$3\%$ for CG-SA; at $N{=}12$ on Topology~2, the corresponding values are about~$1\%$ and~$6\%$.

Taken together, these results show that the hybrid classical--quantum column-generation workflow achieves near-optimal performance once combined with the considered classical reinforcement mechanisms. More importantly, they show that the benefit of the quantum pricing routine is not limited to the raw quality of the sampled solutions: the candidate columns produced by CG-Q are structurally better suited to subsequent classical refinement than those produced by CG-SA.

\section{Conclusion and Perspectives}
\label{sec:conclusion}

In this paper, we addressed the Entanglement Routing in Quantum Networks (ERQN) problem, a fidelity-constrained unsplittable multicommodity flow problem arising in Quantum Information Networks. To cope with its combinatorial complexity, we proposed a hybrid classical--quantum column generation framework that decomposes the routing process into two coordinated tasks: a classical route-selection stage and a route-generation stage. The latter can be reformulated as a constrained shortest-path problem, which remains NP-hard and constitutes a computational bottleneck of the overall framework. To tackle this difficulty, we reformulated the route-generation task as a QUBO problem and solved it through a neutral atom-based quantum optimization workflow combining register embedding and instance-driven pulse shaping.

The proposed approach was evaluated on representative benchmark instances and compared against classical baselines. The results show that the hybrid method achieves near-optimal performance on these instances, with an optimality gap below $1\%$ across all tested sizes when combined with warm-start and post-processing, compared with up to $6\%$ for the classical counterpart based on simulated annealing for route generation. Moreover, even without warm-start, the quantum-assisted route-generation procedure remained below $2\%$, highlighting the structural quality of the routes generated by the neutral atom-based quantum routine. These findings indicate that the benefit of the quantum component does not lie only in the raw quality of the sampled solutions, but also in its ability to produce candidate routes that are particularly effective within the column-generation process and easier to refine classically.

As future work, we plan to extend the network and traffic model to account for the stochastic and time-dependent aspects of entanglement generation and distribution, evaluate the approach on larger and more diverse network instances, and explore stronger route-generation formulations, adaptive penalty calibration, and tighter integration with exact decomposition schemes such as branch-and-price.

\appendix

\section{Compact Arc-Based ILP Formulation}
\label{app:compact_ilp}

This formulation is used for benchmarking with a state-of-the-art ILP solver. We introduce the binary variables
\begin{equation}
y_k =
\begin{cases}
1, & \text{if request }k\text{ is admitted},\\
0, & \text{otherwise},
\end{cases}
\end{equation}
and
\begin{equation}
x_{u,v}^k =
\begin{cases}
1, & \text{if request }k\text{ uses arc }(u,v)\in A,\\
0, & \text{otherwise}.
\end{cases}
\end{equation}
We also use continuous Miller--Tucker--Zemlin (MTZ) order variables $u_u^k\in[0,|V|-1]$.

The compact ILP formulation is then given by
\begin{align}
\max_{x,y}\quad
& \sum_{k\in K}y_k
\label{obj:compact}\\
\text{s.t.}\quad
& \sum_{v:(u,v)\in A}x_{u,v}^k
-\sum_{w:(w,u)\in A}x_{w,u}^k
\notag\\
&\quad=
\begin{cases}
y_k,  & u=s_k,\\
-y_k, & u=t_k,\\
0,    & \text{otherwise},
\end{cases}
\notag\\[-1mm]
& \hspace{8mm}\forall k\in K,\quad
  \forall u\in V,
\label{cons:flow}\\
& \sum_{k\in K}
d_k\left(x_{u,v}^k+x_{v,u}^k\right)
\notag\\
&\quad\le C_{\{u,v\}},
\quad \forall\{u,v\}\in E,
\label{cons:capacity}\\
& \sum_{(u,v)\in A}
\bigl(-\ln\pi_{u,v}\bigr)x_{u,v}^k
\notag\\
&\quad
-\ln\eta
\left(
\sum_{(u,v)\in A}x_{u,v}^k-y_k
\right)
\notag\\
&\quad\le
-\ln\left(\frac{4F_k-1}{3}\right),
\quad \forall k\in K,
\label{cons:fidelity}\\
& u_{s_k}^k=0,
\quad \forall k\in K,
\label{cons:mtz-root}\\
& 1\le u_u^k\le |V|-1,
\notag\\[-1mm]
& \hspace{8mm}
\forall k\in K,\quad
\forall u\in V\setminus\{s_k\},
\label{cons:mtz-bounds}\\
& u_u^k-u_v^k+|V|x_{u,v}^k
\le |V|-1,
\notag\\[-1mm]
& \hspace{8mm}
\forall k\in K,\quad
\forall(u,v)\in A:\ v\neq s_k,
\label{cons:mtz-core}\\
& x_{u,v}^k\in\{0,1\},
\quad y_k\in\{0,1\},
\notag\\[-1mm]
& \hspace{8mm}
\forall k\in K,\quad
\forall(u,v)\in A,
\label{cons:domain-binary}\\
& u_u^k\in[0,|V|-1],
\notag\\[-1mm]
& \hspace{8mm}
\forall k\in K,\quad
\forall u\in V.
\label{cons:domain-order}
\end{align}

The objective~\eqref{obj:compact} maximizes the number of admitted requests. Constraints~\eqref{cons:flow} enforce source-to-destination flow conservation for each request. Constraints~\eqref{cons:capacity} impose edge-capacity limits, while constraints~\eqref{cons:fidelity} enforce the fidelity requirement. Finally, constraints~\eqref{cons:mtz-root}--\eqref{cons:mtz-core} are Miller--Tucker--Zemlin subtour-elimination constraints, which prevent cyclic solutions and ensure that each admitted request is routed along a simple path.
\nocite{*}

\bibliography{apssamp}

\end{document}